\documentclass[12pt]{article}
\usepackage{graphicx}
\usepackage{epsfig}
\usepackage{epstopdf}
\usepackage{url}
\DeclareGraphicsExtensions{.pdf,.eps,.png,.jpg,.mps}

\usepackage{cite}

\def\lsim{\mathrel{\rlap{\lower3pt\hbox{\hskip0pt$\sim$}}
     \raise1pt\hbox{$<$}}}         
\def\gsim{\mathrel{\rlap{\lower4pt\hbox{\hskip1pt$\sim$}}
     \raise1pt\hbox{$>$}}}         

\usepackage{amsmath}
\usepackage{amsfonts}

\begin{document}
\begin{titlepage}

\centerline{\Large \bf Betas, Benchmarks and Beating the Market}
\medskip

\centerline{Zura Kakushadze$^\S$$^\dag$\footnote{\, Zura Kakushadze, Ph.D., is the President and CEO of Quantigic$^\circledR$ Solutions LLC,
and a Full Professor at Free University of Tbilisi. Email: zura@quantigic.com} and Willie Yu$^\sharp$\footnote{\, Willie Yu, Ph.D., is a Research Fellow at Duke-NUS Medical School. Email: willie.yu@duke-nus.edu.sg}}
\bigskip

\centerline{\em $^\S$ Quantigic$^\circledR$ Solutions LLC}
\centerline{\em 1127 High Ridge Road \#135, Stamford, CT 06905\,\,\footnote{\, DISCLAIMER: This address is used by the corresponding author for no
purpose other than to indicate his professional affiliation as is customary in
publications. In particular, the contents of this paper
are not intended as an investment, legal, tax or any other such advice,
and in no way represent views of Quantigic$^\circledR$ Solutions LLC,
the website \url{www.quantigic.com} or any of their other affiliates.
}}
\centerline{\em $^\dag$ Free University of Tbilisi, Business School \& School of Physics}
\centerline{\em 240, David Agmashenebeli Alley, Tbilisi, 0159, Georgia}
\centerline{\em $^\sharp$ Centre for Computational Biology, Duke-NUS Medical School}
\centerline{\em 8 College Road, Singapore 169857}
\medskip
\centerline{(May 30, 2018)}

\bigskip
\medskip

\begin{abstract}
{}We give an explicit formulaic algorithm and source code for building long-only benchmark portfolios and then using these benchmarks in long-only market outperformance strategies. The benchmarks (or the corresponding betas) do not involve any principal components, nor do they require iterations. Instead, we use a multifactor risk model (which utilizes multilevel industry classification or clustering) specifically tailored to long-only benchmark portfolios to compute their weights, which are explicitly positive in our construction.
\end{abstract}
\medskip
\end{titlepage}

\newpage
\section{Introduction and Summary}

{}Diversified long-only portfolios consisting of many stocks are invariably exposed to broad market movements. So, is there an ``optimal" way of constructing such long-only portfolios? Here we can principally distinguish two rather different cases.

{}In the first case, we have no detailed expectations about individual stock returns. I.e., we are trying to construct a long-only portfolio oblivious to any trading signals (or alphas). We can think of such a portfolio as a benchmark. One possibility is to use off-the-shelf (market cap weighted) broad market portfolios such as S\&P 500 or Russell 3000. Another simple approach is to use a minimum-variance portfolio, whose weights $w_i$ ($i=1,\dots,N$ labels $N$ stocks in our universe), up to an overall normalization, are given by
\begin{equation}\label{min-var}
 w_i \propto \sum_{j=1}^N C^{-1}_{ij}
\end{equation}
There are several issues with this. First, if $C_{ij}$ is a sample covariance matrix based on a time-series of historical stock returns, in many cases it is singular as there are not enough observations in the time-series. Second, even if it is nonsingular, the off-diagonal elements of $C_{ij}$ (more, precisely, pair-wise correlations) are highly unstable out-of-sample, and thus so are the benchmark weights $w_i$. Third, the betas $\beta_i$ \cite{Sharpe63} of the individual stocks w.r.t. to this portfolio are all equal 1 (up to an overall normalization factor), so the correlations of the stocks with the benchmark end up being inversely proportional to the historical standard deviations $\sigma_i$ of their returns, which is not what we expect for a broad market benchmark.\footnote{\, It has been appreciated for decades that betas are highly unstable \cite{Fabozzi}.} Fourth, generally, some weights (\ref{min-var}), even if computable, are negative unless we construct a minimum-variance portfolio subject to lower bounds on the weights, thereby either excluding stocks that would otherwise have negative weights\footnote{\, Actually, because of the off-diagonal elements in $C_{ij}$, the stocks that are excluded are not exactly the same as those with $w_i\leq 0$ in (\ref{min-var}).} (thereby, among other things, diminishing diversification and distorting the remainder of the weights), or assigning some ad hoc minimum weights to such stocks (and still distorting the other weights). Some of these issues can be attempted to be circumvented by using (up to an overall normalization factor) the first principal component of $C_{ij}$ as the weights (and, consequently, the betas).\footnote{\, See, e.g.,
\cite{Avellaneda}, \cite{Connor}, \cite{Geweke}, \cite{Trzcinka}.} However, the issues with negative weights and out-of-sample instability (the latter perhaps to a lesser degree) still persist.

{}In the second case, assume that we can forecast non-random expected returns $E_i$ for individual stocks. We can try to construct a long-only portfolio that outperforms a benchmark portfolio. Mean-variance optimization \cite{Markowitz} would give
\begin{equation}
 w_i \propto \sum_{j=1}^N C^{-1}_{ij}~E_j
\end{equation}
This suffers from most of the issues with (\ref{min-var}). One way around this is to: i) first build a benchmark long-only portfolio (which is oblivious to $E_i$); and ii) then build a long-short {\em dollar-neutral} portfolio on top of it based on $E_i$ such that the combined portfolio is still long-only and well-diversified. The dollar-neutral portfolio can be built using standard optimization techniques by employing a well-built, stable multifactor risk model instead of the sample covariance matrix $C_{ij}$. We emphasize that this construction is not simply a ``trend-following" or ``market-timing" strategy. Instead, this is a systematic way of constructing a long-only portfolio as a combination of a long-only ``passive" benchmark\footnote{\, This ``passive" benchmark can be thought of as an index. See, e.g., \cite{Lo2016} for an overview.}  and an ``actively managed" dollar-neutral  portfolio (e.g., based on statistical arbitrage), and the latter is expected to produce positive returns on its own. However, in the context of long-only portfolios, this dollar-neutral strategy is more constrained, as the net positions must be all long. So, this is a way of generating excess returns above that of the ``passive" benchmark, the price being the risk associated with the dollar-neutral strategy (no free lunch).

{}In this paper we discuss a systematic approach to the above ``program". First, we give an explicit formulaic algorithm for constructing a benchmark portfolio given a set of betas, which can be chosen to have various desirable properties. The resultant benchmark weights are positive by construction. This is achieved by building a multifactor risk model $\Gamma_{ij}$ and using it instead of the sample covariance matrix $C_{ij}$ specifically for constructing a long-only benchmark. This $\Gamma_{ij}$ is carefully built based on a binary multilevel industry classification (or some other clustering scheme). It does not involve any principal components, and there are no iterations (due to, e.g., lower bounds) required to obtain $w_i$. In fact, they are given by a simple but nontrivial (and arguably elegant) formula, which is one of the main results of this paper. We give a detailed construction of $\Gamma_{ij}$ and the benchmark weights $w_i$ in Section \ref{sec.3} after reviewing some generalities pertaining to betas in Section \ref{sec.2}. In a nutshell, the weights $w_i$ are expressed via a product of simple algebraic quantities built from specific variances at each level in the industry classification (clustering scheme). These specific variances carry nontrivial information about the underlying stock returns. The source code for computing $\Gamma_{ij}$ and $w_i$ is given in Appendix \ref{app.A}.\footnote{\, The source code given in Appendix \ref{app.A} hereof is not written to be ``fancy" or optimized for speed or in any other way. Its sole purpose is to illustrate the algorithms described in the main text in a simple-to-understand fashion. Some important legalese is relegated to Appendix \ref{app.B}.}

{}In Section \ref{sec.4} we then discuss the outperformance strategy based on overlaying an ``actively managed" dollar-neutral strategy on top of a ``passive" benchmark. Given the expected returns $E_i$, the dollar-neutral portfolio can be constructed using standard optimization techniques with bounds. However, the nontrivial part is that the multifactor risk model $\Gamma^\prime_{ij}$ used in this optimization is {\em different} from $\Gamma_{ij}$ used in constructing the benchmark. Further, this strategy is similar to the S\&P 500 outperformance strategy, where a long position in the S\&P 500 stocks (with dynamic weights) is offset by a short S\&P futures position.\footnote{\, Here: i) we have no futures, and ii) the benchmark is not S\&P 500 based (albeit, it can be).} Section \ref{sec.5} briefly concludes.

\newpage
\section{Betas}\label{sec.2}

{}Consider a universe of $N$ stocks labeled by $i = 1,\dots,N$. Let $R_{is}$ be a time-series of stock returns (e.g., daily, weekly, monthly, etc., close-to-close returns). Here the index $s=1,\dots,T$ labels trading days on which these returns are computed ($s=1$ labels the most recent date).\footnote{\, Here $R_{is}$ can be raw returns or excess returns w.r.t. a risk-free rate, depending on the context.} Let $F_s$ be another time-series of returns, which can but need not correspond to a portfolio of stocks, such as an index. A priori it can be any time-series of returns (e.g., on a commodity, bond, currency, etc.). For our purposes, we can think of $F_s$ as the returns of a benchmark portfolio (see below).

{}Stock betas $\beta_i$ can be thought of as measuring how closely each stock ``follows" the movement of the benchmark. Betas can be formally defined via a {\em serial} regression with the intercept (and unit weights) of the returns $R_{is}$ over $F_s$:
\begin{equation}
 R_{is} = \alpha_i~I_s + \beta_i~F_s + \epsilon_{is}
\end{equation}
Here: $\alpha_i$ is the regression coefficient of the intercept $I_s \equiv 1$; $\beta_i$ is the coefficient of $F_s$; and $\epsilon_{is}$ are the regression residuals. The regression coefficients are given by
\begin{eqnarray}
 &&\alpha_i = {\overline R}_i - \beta_i~{\overline F}\\
 &&\beta_i = {\sum_{s=1}^T {\widetilde R}_{is}~{\widetilde F}_s \over{\sum_{s=1}^T {\widetilde F}^2_s}}
\end{eqnarray}
where for a time-series $A_s$
\begin{eqnarray}
 &&{\overline A}_s = {1\over T}~\sum_{s=1}^T A_s\\
 &&{\widetilde A}_s = A_s - {\overline A}_s
\end{eqnarray}
So, the betas are given by
\begin{equation}\label{def.beta}
 \beta_i = {\mbox{Cov}(R_{is}, F_s)\over{\mbox{Cov}(F_s, F_s)}}
\end{equation}
where $\mbox{Cov}(\cdot,\cdot)$ denotes a serial covariance. If $F_s$ are the returns of a portfolio of the stocks in our universe with weights $w_i$, i.e., if
\begin{equation}
 F_s = \sum_{i=1}^N w_i~R_{is}
\end{equation}
then we have (here $C_{ij}$ is the sample covariance matrix of the stock returns $R_{is}$)
\begin{eqnarray}\label{betas}
 && \beta_i = {1\over \sigma_F^2}~\sum_{j=1}^N C_{ij}~w_j\\
 &&\sigma_F^2 = \sum_{i,j=1}^N C_{ij}~w_i~w_j\\
 &&C_{ij} = \mbox{Cov}(R_{is}, R_{js})
\end{eqnarray}

\section{Benchmarks}\label{sec.3}

{}Suppose we now turn the tables and attempt to construct the benchmark portfolio -- i.e., determine the weights $w_i$ -- given the betas $\beta_i$. This can be done by {\em formally} solving (\ref{betas}) via (in defining ${\widetilde C}_{ij}$ we assume that all $\beta_i\neq 0$):
\begin{eqnarray}\label{weights}
 &&w_i = \sigma_F^2~\sum_{j=1}^N C^{-1}_{ij}~\beta_j = {{\sigma_F^2}\over\beta_i} \sum_{j=1}^N {\widetilde C}^{-1}_{ij} \\
 &&\sigma_F^2 = \left[\sum_{i,j=1}^N C^{-1}_{ij}~\beta_i~\beta_j\right]^{-1} = \left[\sum_{i,j=1}^N {\widetilde C}^{-1}_{ij}\right]^{-1}\label{vol}\\
 &&{\widetilde C}_{ij} = {1\over\beta_i\beta_j}~C_{ij}
\end{eqnarray}
where $C_{ij}^{-1}$ is the matrix inverse to $C_{ij}$. However, in practice there are some issues with this formal solution. First, if $T < N+1$, the sample covariance matrix $C_{ij}$ is singular (i.e., not invertible). Second, unless $T \gg N$, which is seldom (if ever) the case in practice, the off-diagonal elements (in particular, the correlations -- the diagonal elements are relatively stable) are highly unstable out-of-sample rendering $C_{ij}$ essentially useless (unpredictive out-of-sample). Third, even if all $\beta_i$ are positive (e.g., $\beta_i\equiv 1$), assuming $C_{ij}$ is invertible, the weights given by (\ref{weights}) are not necessarily all positive due to nontrivial correlations between stocks. In many applications the benchmark is a long-only portfolio, which requires the weights $w_i$ to be nonnegative.

\subsection{Principal Components}\label{sub.pc}

{}For a moment, let us ignore the aforesaid issues with the sample covariance matrix $C_{ij}$ and assume that it somehow -- magically -- is invertible and stable out-of-sample.\footnote{\, Strictly speaking, invertibility is not required here, but this is not critical.} Then there is a simple ``solution" to constructing a benchmark portfolio.\footnote{\, However, as we discuss in a moment, this ``solution" is intrinsically flawed.} Thus, it is tempting to take $\beta_i = \gamma V^{(1)}_i$, where $\gamma > 0$ is an overall normalization coefficient, and $V^{(1)}_i$ is the first principal component of $C_{ij}$:
\begin{eqnarray}
 &&\sum_{j=1}^N C_{ij}~V^{(a)}_j = \lambda^{(a)}~V^{(a)}_i,~~~a=1,\dots,N\\
 &&\sum_{i=1}^N V^{(a)}_i V^{(b)}_i = \delta_{ab}\label{pc.norm}\\
 &&\sum_{a=1}^N V^{(a)}_i V^{(a)}_j = \delta_{ij}
\end{eqnarray}
Here $V^{(a)}_i$ denotes the $a$-th principal component of $C_{ij}$ corresponding to the eigenvalue $\lambda^{(a)}$, where the eigenvalues are organized in the descending order: $\lambda^{(1)} > \lambda^{(2)} > \dots > \lambda^{(N)}$.\footnote{\, In theory some eigenvalues can be degenerate, but in practice positive eigenvalues are not (and here we are assuming there are no null eigenvalues). This is not critical here as we are interested in the first principal component, whose eigenvalue can be safely assumed to be non-degenerate.} Then the portfolio weights are given by
\begin{equation}\label{prin.comp.1}
 w_i = V^{(1)}_i / \gamma
\end{equation}
More generally, we can take
\begin{eqnarray}
 &&\beta_i = \gamma_i~U^{(1)}_i\label{beta.gamma}\\
 &&w_i = U^{(1)}_i / \gamma_i\\
 &&\sum_{j=1}^N D_{ij}~U^{(a)}_j = {\widetilde\lambda}^{(a)}~U^{(a)}_i,~~~a=1,\dots,N\\
 &&D_{ij} = {1\over\gamma_i\gamma_j}~C_{ij}
\end{eqnarray}
Here $\gamma_i > 0$ is some $N$-vector, and $U^{(a)}_i$ are principal components of the matrix $D_{ij}$.

{}Let us start with (\ref{prin.comp.1}). One immediate issue with this ``solution" is that this portfolio is highly skewed, to wit, it is over-weighted with highly volatile stocks. This is because: i) the stock volatilities $\sigma_i = \sqrt{C_{ii}}$ have a skewed (roughly log-normal) distribution with a long tail for higher values of $\sigma_i$; and ii) the elements of the first principal component roughly scale as $V^{(1)}_i \propto \sigma_i$. To see this, consider a simple model where all stocks have uniform pair-wise correlations equal $\rho$:\footnote{\, I.e., the sample correlation matrix $\Psi_{ij} = C_{ij}/\sigma_i\sigma_j$ is replaced by a 1-factor model (see below), where $\rho$ can be taken as the average pair-wise correlation ${\overline\rho} = \sum_{i,j=1;~i\neq j}^N \Psi_{ij}/N(N-1)$. This is an oversimplified model, but (as we discuss below) it correctly captures the point we make here.}
\begin{equation}\label{uniform}
 C_{ij} = \sigma_i\sigma_j\left[(1-\rho)\delta_{ij} + \rho\nu_i\nu_j\right]
\end{equation}
where $\nu_i\equiv 1$ is the unit $N$-vector. Straightforward algebra leads to the following eigenvalue equation:
\begin{equation}\label{eigen.eq}
 \rho~\sum_{i=1}^N {\sigma_i^2\over{\lambda - (1-\rho)\sigma_i^2}} = 1
\end{equation}
whose $N$ solutions are the eigenvalues $\lambda = \lambda^{(a)}$, $a=1,\dots,N$. The eigenvectors are given by:
\begin{equation}
 V^{(a)}_i = {\eta^{(a)}~\sigma_i\over{\lambda^{(a)} - (1-\rho)\sigma_i^2}}
\end{equation}
where $\eta^{(a)}$ are overall normalization factors (fixed via (\ref{pc.norm})). We can solve (\ref{eigen.eq}) iteratively via
\begin{equation}
 \lambda = \rho~\sum_{i=1}^N {\sigma_i^2\over{1 - (1-\rho)\sigma_i^2/\lambda}}
\end{equation}
assuming $\lambda \gg \rho~\sigma_i^2$ for all $i$. In the zeroth approximation we have\footnote{\, The actual $\lambda^{(1)}$ is even higher and the actual $V^{(1)}_i$ are even more skewed for large $\sigma_i$, which further aids the point we make here. At any rate, in practice the zeroth approximation is pretty good. E.g., for the dataset discussed in Subsection 2.6 of \cite{Fano}, which consists of 21-day historical volatilities based on daily close-to-close
returns for 3810 U.S. stocks, we have the following statistics for the ratio $\sigma_i^2 / \sum_{j=1}^N \sigma_j^2$: Min = $1.70 \times 10^{-7}$, 1st Quartile =  $3.48 \times 10^{-5}$, Median = $7.79 \times 10^{-5}$, Mean = $2.63 \times 10^{-4}$, 3rd Quartile = $2.00 \times 10^{-4}$, Max = $4.39 \times 10^{-2}$. Therefore, the zeroth approximation (\ref{zeroth.appr}) is justified assuming $\rho$ is not too small, which is the case considering that the 21-day historical average pair-wise correlation for the aforesaid dataset ${\overline\rho} = 0.10$. So, if we set $\rho={\overline\rho}$, then for $\kappa_i = (1-\rho)\sigma_i^2 / \rho\sum_{j=1}^N \sigma_j^2$ we have that $\mbox{max}(\kappa_i)\approx 0.4$ and for most stocks $\kappa_i \ll 1$. Further, a relatively small number of stocks for which $\kappa_i$ is not small have even larger contributions to $\lambda^{(1)}$ and $V^{(1)}_i$ compared with the aforesaid zeroth approximation, which further exacerbates the skewness of $V^{(1)}_i$ (and thus $w_i$) for the large $\sigma_i$ stocks (see below).\label{fn.data}}
\begin{eqnarray}
 &&\lambda^{(1)} \approx \rho~\sum_{i=1}^N\sigma_i^2 \label{zeroth.appr}\\
 && V^{(1)}_i \approx {\sigma_i\over\sqrt{\sum_{i=1}^N\sigma_i^2}}
\end{eqnarray}
It is now evident that the portfolio given by the weights (\ref{prin.comp.1}) is indeed skewed and overloaded with high volatility stocks. This conclusion persists even if we do not assume uniform correlations as in (\ref{uniform}). Thus, straightforward algebra will convince the reader that this is the case for, e.g., the following model:
\begin{eqnarray}
 &&C_{ij} = \sigma_i\sigma_j{\widetilde \Psi}_{ij}\label{1-factor.stat}\\
 &&{\widetilde\Psi}_{ij} = {\widetilde \xi}_i^2~\delta_{ij} + {\widetilde\lambda}^{(1)}~U^{(1)}_i~U^{(1)}_j\\
 &&{\widetilde \xi}_i^2 = 1 - {\widetilde\lambda}^{(1)}~[U^{(1)}_i]^2
\end{eqnarray}
where $U^{(1)}_i$ is the first principal component of the sample correlation matrix $\Psi_{ij} = C_{ij}/\sigma_i\sigma_j$, and ${\widetilde\lambda}^{(1)}$ is the corresponding eigenvalue.\footnote{\, In (\ref{1-factor.stat}) the sample correlation matrix $\Psi_{ij}$ is replaced by a 1-factor statistical risk model ${\widetilde\Psi}_{ij}$ based on the first principal component $U^{(1)}_i$ of $\Psi_{ij}$. See \cite{StatRM} for details.\label{fn.SRM}} Note that $0 < {\widetilde \xi}_i^2 < 1$. Now we no longer have uniform correlations, and (in the zeroth approximation) we have $V^{(1)}_i \propto \sigma_i~U^{(1)}_i$, so $V^{(1)}_i$ are skewed for large $\sigma_i$ (as $U^{(1)}_i$ are independent of $\sigma_i$ and are not skewed; in the zeroth approximation $U^{(1)}_i\approx 1/\sqrt{N}$). So the problem persists.

{}A simple ``fix" is to take skewed betas to begin with, e.g., if we take $\gamma_i = \sigma_i$ in (\ref{beta.gamma}), then we have $w_i = U^{(1)}_i/\sigma_i$, where $U^{(1)}_i$ is the first principal component of the sample correlation matrix $\Psi_{ij}$. Now $w_i$ are skewed but in the opposite direction -- the weights for volatile stocks are suppressed, which is a welcome feature. This would seem to solve the problem of constructing an acceptable benchmark. However, there are other issues we have been postponing dealing with. First, generally, some elements of $U^{(1)}_i$ are negative.\footnote{\, E.g., in the dataset mentioned in fn. \ref{fn.data}, out of the 3810 elements, 575 (or over 15\%) of $U^{(1)}_i$ and 668 elements of $V^{(1)}_i$ are negative. This is a sizable number.} Since for long-only benchmarks we wish to have $w_i > 0$, this poses an issue. One way to circumvent this is not to use (\ref{beta.gamma}) (with $\gamma_i = \sigma_i$), but instead to proceed as follows. First, instead of using the sample covariance matrix $C_{ij}$, we model it via the 1-factor statistical risk model (\ref{1-factor.stat}) (see fn. \ref{fn.SRM}). The inverse of $C_{ij}$ is given by
\begin{eqnarray}
 &&C_{ij}^{-1} = {1\over \sigma_i\sigma_j}~{\widetilde\Psi}_{ij}^{-1} \\
 &&{\widetilde\Psi}_{ij}^{-1} = {1\over {\widetilde\xi}_i^2}~\delta_{ij} - q~{U^{(1)}_i \over {\widetilde\xi}_i^2}~ {U^{(1)}_j \over {\widetilde\xi}_j^2}\\
 &&q = \left[{1/{\widetilde\lambda}^{(1)}} + \sum_{k=1}^N {[U^{(1)}_k]^2 /{\widetilde\xi}_k^2}\right]^{-1}
\end{eqnarray}
Let\footnote{\, In practice, generically, no element of $U^{(1)}_i$ is expected to be exactly 0.} $H_+ =\{i| U^{(1)}_i \geq 0\}$ and $H_+ =\{i| U^{(1)}_i < 0\}$. Then we can set (here $\gamma$ is an overall normalization constant, while $\kappa$ is a parameter to be fixed)
\begin{eqnarray}
 &&\beta_i = \gamma~\sigma_i~U^{(1)}_i,~~~i\in H_+\\
 &&\beta_i = \kappa~\gamma~\sigma_i~U^{(1)}_i,~~~i\in H_-\\
 &&w_i = {{\widetilde q}~ U^{(1)}_i\over \gamma\sigma_i{\widetilde\xi}_i^2} \left[{1/{\widetilde\lambda}^{(1)}} + \left(1-\kappa\right)G_- \right],~~~i\in H_+\\
 &&w_i = {{\widetilde q}~ U^{(1)}_i\over \gamma\sigma_i{\widetilde\xi}_i^2} \left[{\kappa/{\widetilde\lambda}^{(1)}} - \left(1-\kappa\right) G_+\right],~~~i\in H_-\\
 &&{\widetilde q} = \left[{(G_+ + \kappa^2~G_-)/{\widetilde\lambda}^{(1)}} + \left(1-\kappa\right)^2 G_+~G_-\right]^{-1}\\
 &&G_+ = \sum_{k\in H_+} {[U^{(1)}_k]^2/ {\widetilde\xi}_k^2}\\
 &&G_- = \sum_{k\in H_-} {[U^{(1)}_k]^2/ {\widetilde\xi}_k^2}
\end{eqnarray}
It then follows that for $i\in H_-$ we have: $w_i < 0$ for $\kappa_* < \kappa \leq 1$; $w_i = 0$ for $\kappa = \kappa_*$; and $w_i > 0$ for $\kappa < \kappa_*$. Here
\begin{equation}\label{kappa.star}
 \kappa_* = {G_+ \over{{1/{\widetilde\lambda}^{(1)}} + G_+}}
\end{equation}
Now: i) $N_+ = |H_+|$ is sizably larger than $N_- = |H_-|$ (so $G_+ \gsim 1$); and ii) $1/{\widetilde\lambda}^{(1)}\ll 1$, so the corresponding contribution to $\kappa_*$ in (\ref{kappa.star}) is subleading. As a result, $\kappa_*\approx 1 - 1/G_+{\widetilde\lambda}^{(1)}$ is very close to 1. Also, $G_+ \gg G_-$. Na\"{\i}vely, it might appear reasonable to assume that $G_+/G_-\sim N_+/N_-$. However, typically, $G_+/G_-$ is sizably larger than $N_+/N_-$. The additional skew is due to the fact that the absolute value of the average negative pair-wise correlation is sizably lower than the average positive pair-wise correlation,\footnote{\, Thus, for the dataset mentioned in fn. \ref{fn.data}, the median (mean) correlation in $H_+$ is $0.214$ ($0.239$), while in $H_-$ it is $-0.136$ ($-0.165$). The median (mean) $[U^{(1)}_i]^2$ in $H_+$ is $2.40 \times 10^{-4}$ ($2.99 \times 10^{-4}$), while in $H_-$ it is $2.22 \times 10^{-5}$ ($5.82 \times 10^{-5}$). The median (mean) ${\widetilde\xi}_i^2$ in $H_+$ is $0.838$ ($0.799$), while in $H_-$ it is $0.985$ ($0.961$). Also, $G_+ = 1.637$, $G_- = 0.03919$, and ${\widetilde\lambda}^{(1)} = 674.35$.} which implies that average $[U^{(1)}_i]^2$ for $i\in H_-$ is sizably lower than average $[U^{(1)}_i]^2$ for $i\in H_+$, which in turn implies that on average ${\widetilde\xi}_i^2$ are closer to 1 for $i\in H_-$ than for $i\in H_+$. So, let $\kappa = 1 - \zeta/G_+{\widetilde\lambda}^{(1)}$. For $\zeta=1$ we have $\kappa \approx \kappa_*$ (up to subleading corrections). For $\zeta > 1$ we have $w_i > 0$ for $i\in H_-$. However, if we take $\zeta \gg 1$, then on average the weights $w_i$ for $i\in H_-$ will be much higher than for $i\in H_+$. This implies that $\zeta\sim 1$. Then we have (up to subleading corrections)
\begin{eqnarray}
 &&w_i \approx {U^{(1)}_i\over \gamma\sigma_i{\widetilde\xi}_i^2 G_+},~~~i\in H_+\\
 &&w_i \approx -{U^{(1)}_i\over \gamma\sigma_i{\widetilde\xi}_i^2 G_+} \left[\zeta - 1\right],~~~i\in H_-
\end{eqnarray}
So, we can determine the value of $\zeta = \zeta_*$ such that the mean (or median)\footnote{\, For the dataset mentioned in fn. \ref{fn.data}, the value of $\zeta_*$ computed based on the median (mean) is $6.988$ ($6.181$). This reflects the asymmetry between the stocks in $H_+$ and $H_-$ mentioned above.} of $w_i$ is the same in $H_+$ and $H_-$. This way stocks from these two sets will on average be weighted more or less equally in the benchmark, with all $w_i>0$ (and $\beta_i < 0$ in $H_-$).

{}Here one may argue that the solution with $\zeta = 2$, where $w_i \approx {|U^{(1)}_i|/ \gamma\sigma_i{\widetilde\xi}_i^2 G_+}$, is more natural (notwithstanding that on average the weights $w_i$, $i\in H_-$, are somewhat smaller than the weights $w_i$, $i\in H_+$). However, there is no magic bullet here for fixing $\zeta$. In fact, the parametrization using a single parameter, $\zeta$, is just one possibility out of myriad others. Now, as to $\zeta$, it can be chosen to be pretty much anywhere between (somewhat above) 1 and (somewhat above) $\zeta_*$ defined above.

{}So, problem solved? Well, not quite. The fact that tiny (of order $1/G_+{\widetilde\lambda}^{(1)}$ on a relative basis) changes in $\beta_i$ produce of order 1 changes in the benchmark weights $w_i$ should sound an alarm. The culprit here is that, as already mentioned, in the zeroth approximation we have $U^{(1)}_i\approx 1/\sqrt{N}$. This is the so-called ``market mode" -- the first principal component of $\Psi_{ij}$ corresponds to the overall movement of the broad market (see, e.g., \cite{CFM}). The deviations from $U^{(1)}_i\approx 1/\sqrt{N}$ correspond to different stocks having different $\beta_i/\sigma_i$. However, in the simple 1-factor model (\ref{1-factor.stat}), very different looking benchmarks have almost the same betas. To see this, consider the weights $w_i = \omega_i/\sigma_i$, where $0 < \omega_i \sim 1$. We then have
\begin{equation}
 \beta_i = {\sigma_i\over\sigma_F^2}\left[{\widetilde\xi}_i^2~\omega_i + {\widetilde\lambda}^{(1)}~U^{(1)}_i~\sum_{j=1}^N U^{(1)}_j~\omega_j\right]
\end{equation}
Considering that ${\widetilde\xi}_i^2\sim 1$ and ${\widetilde\lambda}^{(1)} \gg 1$, the first term due to the specific risk is subleading and the second term is dominant, so we have
\begin{equation}
 \beta_i = {\sigma_i\over\sigma_F^2} ~{\widetilde\lambda}^{(1)}~U^{(1)}_i~\sum_{j=1}^N U^{(1)}_j~\omega_j\left[1 + {\cal{O}}(1/{\widetilde\lambda}^{(1)})\right]
\end{equation}
So, up to small ${\cal{O}}(1/{\widetilde\lambda}^{(1)})$ corrections, $\beta_i$ are simply proportional to $\sigma_i U^{(1)}_i$, irrespective of the individual values of $\omega_i$. Therefore, all that gymnastics for constructing $w_i>0$ is just a mirage and an exercise in self-deception. Is there a way around this?

{}To get betas to substantially deviate from $\beta_i\propto \sigma_i U^{(1)}_i$, we can try to include higher principal components via a $K$-factor statistical risk model:
\begin{eqnarray}
 &&C_{ij} = \sigma_i\sigma_j{\widetilde \Psi}_{ij}\label{K-factor.stat}\\
 &&{\widetilde\Psi}_{ij} = {\widetilde \xi}_i^2~\delta_{ij} + \sum_{a=1}^K {\widetilde\lambda}^{(a)}~U^{(a)}_i~U^{(a)}_j\\
 &&{\widetilde \xi}_i^2 = 1 - \sum_{a=1}^K {\widetilde\lambda}^{(a)}~[U^{(a)}_i]^2
\end{eqnarray}
Here $K>1$ can be fixed using eRank (effective rank) of \cite{RV} or some other method -- see \cite{StatRM} for details. A ``technical" issue with this approach is that ensuring $w_i>0$ becomes messier. A more important, conceptual issue is that higher principal components are intrinsically unstable out-of-sample. This instability is then inherited by the benchmark weights $w_i$ and the betas $\beta_i$ constructed using such models. This is not a fruitful direction to pursue.

\subsection{Back to a 1-factor Model}\label{sub.back.1-factor}

{}While we can and will consider multifactor models beyond statistical risk models, we are not done with 1-factor models quite yet. Let us consider a general 1-factor model:
\begin{eqnarray}
 &&C_{ij} = \sigma_i\sigma_j{\widetilde \Psi}_{ij}\label{1-factor.gen}\\
 &&{\widetilde\Psi}_{ij} = {\widetilde \xi}_i^2~\delta_{ij} + \Omega_i~\Omega_j\\
 &&{\widetilde \xi}_i^2 = 1 - \Omega_i^2
\end{eqnarray}
A priori we must only require that $\Omega_i^2 < 1$ and $\Omega_i$ are arbitrary otherwise. If we take $\Omega_i = \beta_i/\sigma_i$, straightforward algebra gives
\begin{eqnarray}\label{w.1-factor.beta}
 &&w_i = \eta~{\beta_i\over\xi_i^2}\\
 &&\xi^2_i = \sigma_i^2~{\widetilde\xi}_i^2\\
 &&\eta^{-1} = \sum_{i=1}^N {\beta_i^2\over\xi_i^2}
\end{eqnarray}
Here $\xi_i^2$ is the specific variance in the covariance matrix $C_{ij}$ (as opposed to ${\widetilde\xi}_i^2$, which is the same quantity in the correlation matrix ${\widetilde\Psi}_{ij}$). So, up to an overall normalization factor $\eta$, $w_i$ are proportional to $\beta_i/\xi_i^2$, which is the same result as what we would have obtained had we assumed -- albeit this may appear strange at first -- that $C_{ij} = \xi_i^2~\delta_{ij}$. Put differently, it is as though we ignore the factor risk altogether and only account for specific (idiosyncratic) risk. However, there is actually nothing strange about this result. Importantly, if all $\beta_i > 0$, then automatically all $w_i > 0$.

\subsection{Interpretation}

{}The weights (\ref{weights}) are nothing but the solution to maximizing the expected Sharpe ratio \cite{Sharpe66}, \cite{Sharpe94}\footnote{\, Here we have no transaction costs, bounds or constraints, so maximizing the Sharpe ratio is equivalent to mean-variance optimization \cite{Markowitz}. See, e.g., \cite{MeanRev}.} $S$ of the benchmark portfolio if we treat $\beta_i$ as the expected returns $E_i$ for our stocks:
\begin{equation}\label{exp.ret}
 E_i = \gamma~\beta_i
\end{equation}
where $\gamma$ is an immaterial (for our purposes here) overall normalization factor. The expected Sharpe ratio is given by
\begin{eqnarray}\label{Sharpe}
 && S = {1\over \sigma_F}~\sum_{i=1}^N E_i~w_i\\
 && \sigma_F^2 = \sum_{i,j=1}^N C_{ij}~w_i~w_j
\end{eqnarray}
Maximizing $S$ w.r.t. $w_i$, we get (\ref{weights}) up to an overall normalization factor. The latter can be fixed by requiring that
\begin{equation}\label{norm.w.beta}
 \sum_{i=1}^N w_i~\beta_i = 1
\end{equation}
which is a consequence of (\ref{weights}) and (\ref{vol}). Now we can understand the issue we encountered above with the ``market mode". Maximizing the Sharpe ratio hedges\footnote{\, Not precisely, but approximately. However, when $N$ is large, for the 1-factor model (e.g., based on the first principal component $U^{(1)}_i$ as in (\ref{1-factor.stat})), the ``mishedge" is suppressed by $1/N$ \cite{StatRM}.} against the broad market going bust (i.e., all or most stocks selling off at the same time). While this is a natural thing to do when constructing dollar-neutral portfolios, it makes no sense to do this when constructing long-only portfolios. Indeed, long only portfolios are exposed to market risk by definition. Hedging against specific risk does make sense, which amounts to dropping the factor risk from $C_{ij}$ (in a 1-factor model) and simply taking $C_{ij} = \xi_i^2~\delta_{ij}$, i.e., treating it as though we only have specific risk. So, the lesson here is that we must eliminate the ``market mode".

\subsection{Multifactor Models}

{}Next, let us discuss a general multifactor model covariance matrix $\Gamma_{ij}$:
\begin{equation}\label{factor}
 \Gamma_{ij} = \xi_i^2~\delta_{ij} + \sum_{A,B=1}^K \Omega_{iA}~\phi_{AB}~\Omega_{jB}
\end{equation}
Here: $\xi_i^2$ is the specific (a.k.a. idiosyncratic) risk; $\Omega_{iA}$, $A=1,\dots,K$, is the factor loadings matrix; and $\phi_{AB}$ is the factor covariance matrix. For our purposes here it will not be important to know how $\Gamma_{ij}$ is constructed.\footnote{\, For a general discussion, see, e.g., \cite{GK}. For an explicit open-source implementation of a general multifactor risk model for equities, see \cite{HetPlus}.} What matters here is that: the number of risk factors $K\ll N$ ($K$ can still be in hundreds); all $\xi_i > 0$; $\phi_{AB}$ is positive-definite (then so is $\Gamma_{ij}$); $\Gamma_{ii} = C_{ii}$ (so the sample variances are matched).

{}Using the inverse of $\Gamma_{ij}$
\begin{eqnarray}
 &&\Gamma_{ij}^{-1} = {1\over \xi_i^2}~\delta_{ij} - \sum_{A,B=1}^K {\Omega_{iA}\over\xi_i^2}~Q^{-1}_{AB}~{\Omega_{jB}\over \xi_j^2}\\
 &&Q_{AB} = \phi^{-1}_{AB} + \sum_{i=1}^N {1\over\xi_i^2}~\Omega_{iA}~\Omega_{iB}
\end{eqnarray}
in lieu of $C_{ij}^{-1}$ in (\ref{weights}) and (\ref{vol}), we have
\begin{eqnarray}\label{ww}
 &&w_i = {\sigma_F^2\over\xi_i^2} \left[\beta_i - \Upsilon_i\right]\\
 &&\Upsilon_i = \sum_{A,B=1}^K \Omega_{iA}~Q^{-1}_{AB}~\Lambda_B \label{Upsilon0}\\
 &&\Lambda_A = \sum_{j=1}^N {\beta_j~\Omega_{jA}\over \xi_j^2}\\
 &&\sigma_F^2 = \left[\Theta - \sum_{A,B=1}^K \Lambda_A~Q^{-1}_{AB}~\Lambda_B\right]^{-1}\\
 &&\Theta = \sum_{j=1}^N {\beta_j^2\over \xi_j^2}
\end{eqnarray}
The weights $w_i$ can be rewritten as follows. Let
\begin{eqnarray}
 &&{\widetilde \Omega}_{iA} = {1\over\xi_i^2}\left[\Omega_{iA} - \beta_i~{\Lambda_A\over \Theta}\right]\\
 &&{\widetilde \Upsilon}_i = \sum_{A,B=1}^K {\widetilde \Omega}_{iA}~Q^{-1}_{AB}~\Lambda_B
\end{eqnarray}
Straightforward algebra results in the following expression for $w_i$:
\begin{equation}\label{w1}
 w_i = {\beta_i\over \Theta~\xi_i^2} - \sigma_F^2~{\widetilde \Upsilon}_i
\end{equation}
Note that (thus, we have (\ref{norm.w.beta}))
\begin{equation}\label{beta-neutral}
 \sum_{i=1}^N\beta_i~{\widetilde\Upsilon}_i = 0
\end{equation}
Therefore, assuming all $\beta_i>0$, the first term in (\ref{w1}) is always positive; however, ${\widetilde\Upsilon}_i$ can be negative (in which case $w_i$ is positive) or positive, in which case $w_i$ can be negative. A deceptively ``simple" way to ensure that all $w_i > 0$ is to take $\Omega_{iA}$ such that they are ``orthogonal" to $\beta_i$
\begin{equation}\label{gen.orth}
 \sum_{j=1}^N {\beta_j~\Omega_{jA}\over \xi_j^2}\equiv 0
\end{equation}
In this case we have $\Lambda_A\equiv 0$ and ${\widetilde\Upsilon}_i \equiv 0$. However, this is not practicable as a priori (i.e., before constructing the full risk model) $\xi_i^2$ are unknown and their dependence on $\Omega_{iA}$ is highly nonlinear (see \cite{HetPlus} for details). On the other hand, it is also impracticable to derive conditions ensuring that $w_i > 0$ (or at least $w_i \geq 0$) for a general multifactor model. Nonetheless, (\ref{gen.orth}) has an important interpretation: it is nothing but the requirement that the risk factors be ``orthogonal" to the ``market mode". In fact, we can turn this around and ask: what happens if we include the ``market mode"? We already know the answer in the case of a 1-factor model: if we take the corresponding factor loading as $\beta_i$, then the factor risk does not affect the weights $w_i$. But for a multifactor model things are trickier.

{}So, let us assume that $\Omega_{i1} = \beta_i$. Then we can always rotate the remaining $\Omega_{iA}$, $A>1$, such that $\Lambda_A \equiv 0$ for $A>1$. (Note that this rotation affects $\phi_{AB}$.) We then have ${\widetilde\Omega}_{i1} \equiv 0$, while ${\widetilde\Omega}_{iA} = \Omega_{iA}/\xi_i^2$ for $A>1$. Therefore, we have
\begin{eqnarray}
 &&{\widetilde \Upsilon}_i = \Theta~ \sum_{A=2}^K {\widetilde \Omega}_{iA}~Q^{-1}_{A1}\\
 &&Q_{A1} = \phi^{-1}_{A1},~~~A > 1
\end{eqnarray}
So, if $\phi_{A1}\equiv 0$ for $A>1$, then for these values of $A$, we also have $\phi^{-1}_{A1}\equiv 0$, $Q_{A1}\equiv 0$ and $Q^{-1}_{A1}\equiv 0$, which would imply that ${\widetilde \Upsilon}_i\equiv 0$. It is therefore the nonzero correlations (i.e., mixing) between the ``market mode" and the other risk factors that make ${\widetilde \Upsilon}_i\neq 0$, which can lead to some negative $w_i$. Barring setting $\phi_{A1}\equiv 0$ for $A>1$ ad hoc (which is impracticable for the same reasons as above), there is no simple way to guarantee that $w_i>0$ if we do not exclude the ``market mode". To be sure, as already mentioned, there is no practicable way to guarantee (\ref{gen.orth}) either. However, our point here is that not excluding the ``market mode" further exacerbates the issue. So, are we at sea? Not quite. We just need to dig deeper.

\subsection{Binary ``Cluster" Factors}

{}Thus, in practice, the factor loadings are not arbitrary but relatively constrained. The columns of the factor loadings matrix $\Omega_{iA}$ typically are based on: i) industry classification (or some other clustering); ii) style factors (e.g., size, value, liquidity, volatility, etc.); and/or iii) principal components. We already discussed principal components and we will return to them a bit later. We will also come back to style factors. Here we focus on ``cluster" based factors, which can be based on a fundamental industry classification or a statistical one \cite{StatIC}.

{}First, let $\sigma_i^2 = C_{ii}$ be the total sample variances computed based on the historical time-series data. We have (see above) $\Gamma_{ii} = \sigma_i^2$. The volatilities $\sigma_i$ have a skewed (roughly log-normal) cross-sectional distribution with a long tail for higher values of $\sigma_i$. This is why in practice, instead of directly modeling $C_{ij}$ via a factor model, it makes a lot more sense to model the sample correlation matrix $\Psi_{ij} = C_{ij} / \sigma_i\sigma_j$, from which the skewness present in $\sigma_i$ has been nicely factored out.\footnote{\, For details, see \cite{Het}, \cite{HetPlus}.} Indeed, the diagonal elements $\Psi_{ii}\equiv 1$, and the off-diagonal ones (i.e., pair-wise correlations) $|\Psi_{ij}| < 1$ ($i\neq j$). What this means in terms of the factor model for $\Gamma_{ij}$ is that
\begin{eqnarray}
 &&\Gamma_{ij} = \sigma_i\sigma_j{\widehat \Gamma}_{ij}\\
 &&{\widehat \Gamma}_{ij} = {\widehat\xi}_i^2~\delta_{ij} + \sum_{A,B=1}^K {\widehat \Omega}_{iA}~\phi_{AB}~{\widehat\Omega}_{jB} \label{Gammahat}\\
 &&{\widehat \xi}_i = \xi_i /\sigma_i\\
 &&{\widehat\Omega}_{iA} = \Omega_{iA} / \sigma_i\\
 &&{\widehat\xi}_i^2 + \sum_{A,B=1}^K {\widehat \Omega}_{iA}~\phi_{AB}~{\widehat\Omega}_{iB}\equiv 1
\end{eqnarray}
The matrix ${\widehat\Omega}_{iA}$ is wholly devoid of any skewness in $\sigma_i$. This simplifies things a lot.

{}Now, let us consider a model where the factor loadings ${\widehat\Omega}_{iA}$ are based on a binary industry classification:
\begin{eqnarray}
 &&{\widehat\Omega}_{iA} = \Omega_i~\delta_{G(i), A}\\
 &&G:\{1,\dots,N\}\mapsto\{1,\dots,K\}
\end{eqnarray}
Here: the $N$-vector $\Omega_i$ a priori is arbitrary; $G$ maps stocks labeled by $i$ ($i=1,\dots,N$) to ``clusters" labeled by $A$ ($A=1,\dots,K$); each cluster contains one and only one stock; $J(A)=\{i|G(i) = A\}$ is the set of stocks that belong to the cluster labeled by $A$; and $N(A) = |J(A)|$ is the number of stocks in said cluster. The clusters can be, e.g., sectors, industries or sub-industries in a binary industry classification.\footnote{\, Such as GICS (Global Industry Classification Standard), BICS (Bloomberg Industry Classification System), SIC (Standard Industrial Classification), etc. In principle, we can also consider quasi-binary classifications where some stocks (conglomerates, whose number typically is relatively small) belong to more than one cluster. We will not do so, nor will it be critical for our purposes.} We then have
\begin{equation}
 {\widehat\xi}_i^2 = 1 - \Omega_i^2~\phi_{G(i),G(i)}
\end{equation}
which implies that (by definition, $\phi_{AB}$ is the factor covariance matrix, so $\phi_{AA} > 0$)
\begin{equation}
 \Omega_i^2 < 1 / \phi_{G(i),G(i)}
\end{equation}
For $\Omega_i$, we basically have three choices. We can take $\Omega_i = 1/\sqrt{N(A)}$, $i\in J(A)$, i.e., uniform within-cluster loadings.\footnote{\, This is the binary risk model construction \cite{HetPlus}.} Another choice is to take $\Omega_i = [U(A)]_i$, $i\in J(A)$, where the $N(A)$-vector $[U(A)]_i$ is the first principal component of the $N(A)\times N(A)$ matrix $[\Psi(A)]_{ij} = \Psi_{ij}$, $i,j\in J(A)$.\footnote{\, This is the heterotic construction \cite{Het}, \cite{HetPlus}.} Finally, we can take $\Omega_i$ to be a style factor.\footnote{\, This is the heterotic CAPM construction \cite{HetPlus}. The ``style" factor here can be related to $\beta_i$ itself (see below).}

{}Once $\Omega_i$ is specified, the factor covariance matrix $\phi_{AB}$ and the specific risks ${\widehat \xi}_i$ can be computed.\footnote{\, More precisely, depending on the historical lookback, $\phi_{AB}$ may be computable as a sample covariance matrix of factor returns or itself may have to be modeled via a factor model covariance matrix as the sample factor covariance matrix may be singular or unstable out-of-sample. However, the diagonal elements $\phi_{AA}$ are always the same as sample variances of the (appropriately defined and normalized) factor returns. See \cite{Het}, \cite{HetPlus}.} So, following our discussion in the 1-factor case, let us take $\Omega_i = \beta_i/\sigma_i$. Then straightforward algebra gives
\begin{eqnarray}
 && w_i = \sigma_F^2~{\beta_i\over\xi_i^2}~\gamma_{G(i)}\label{w.cluster}\\
 && \xi_i^2 = \sigma_i^2~{\widehat \xi}_i^2\\
 && \gamma_A = 1 - \sum_{B=1}^K Q^{-1}_{AB}~\Lambda_B\\
 && \Lambda_A = \sum_{j \in J(A)} {\beta_i^2\over\xi_i^2}\\
 && Q_{AB} = \phi_{AB}^{-1} + \Lambda_A~\delta_{AB}\\
 && \sigma_F^{-2} = \sum_{A=1}^K \Lambda_A~\gamma_A\label{gamma.Lambda}
\end{eqnarray}
Here $\xi_i^2$ is the specific variance in the factor model covariance matrix $\Gamma_{ij}$ (as opposed to ${\widehat\xi}_i^2$, which is the same quantity in the correlation matrix ${\widehat\Gamma}_{ij}$). So, the important lesson from (\ref{w.cluster}) is that the weights $w_i$ within each cluster are computed the same way as in the 1-factor model (i.e., by ignoring the factor risk). The normalization factors $\gamma_{G(i)}$ are uniform within each cluster and only vary from cluster to cluster. They result from optimizing cluster returns across the $K$ clusters (see below). In this regard, they are not guaranteed to be positive. Thus, if we assume that there is no mixing between the clusters, i.e., the pair-wise factor correlations vanish ($\phi_{AB} \equiv 0$, $A\neq B$), then we have $\gamma_A = 1 / (1+ \phi_{AA}\Lambda_A) > 0$. In the presence of mixing we can have some negative $\gamma_A$. However, then all $w_i$ in the corresponding cluster(s) would be negative as well (assuming all $\beta_i>0$). This is an artifact of optimizing cluster returns, which (approximately) hedges against all clusters going bust. As in the case of a 1-factor model, for long-only portfolios it makes no sense to do this. E.g., if clusters are sectors and our broad benchmark contains stocks in all sectors (with each sector itself well-diversified), it makes no sense to hedge against all sectors going bust -- the benchmark is long all sectors by definition. Mathematically, this can be understood as the factor covariance matrix containing its own ``market mode" (corresponding to the overall movement of all sectors), which must be eliminated when computing the benchmark weights $w_i$. This can be seen by modeling $\phi_{AB}$ via a 1-factor model. The story is the same as above, with clusters in the place of stocks.

\subsubsection{Example: 1-factor $\phi_{AB}$}

{}To illustrate the discussion above, for our purposes here it suffices to consider a 1-factor model for the factor covariance matrix:
\begin{equation}\label{1-factor.phi}
 \phi_{AB} = \zeta_A^2~\delta_{AB} + \chi_A\chi_B
\end{equation}
Straightforward algebra then yields
\begin{eqnarray}
 &&Q_{AB}^{-1} = {1\over\nu_A^2}~\delta_{AB} + {1\over\kappa}~{\chi_A\over\nu_A^2\zeta_A^2}~{\chi_B\over\nu_B^2\zeta_B^2}\\
 &&\nu_A^2 = {1\over\zeta_A^2} + \Lambda_A \\
 &&\kappa = 1 + \sum_{A=1}^K {\chi_A^2 \Lambda_A\over{1+\zeta_A^2\Lambda_A}}
\end{eqnarray}
So, we have
\begin{equation}\label{gamma.A}
 \gamma_A =
 {\kappa^{-1}\over{1+\zeta_A^2\Lambda_A}}\left( 1 - \sum_{B=1;~B\neq A}^K \left[{\chi_A\over\chi_B} - 1\right] {\chi^2_B\Lambda_B \over{1+\zeta_B^2\Lambda_B}}\right)
\end{equation}
For generic $\chi_A$ (i.e., $|\chi_A/\chi_B - 1|\sim 1$, $B\neq A$), some $\gamma_A$ can be negative. Consider $A=A_*$, $\chi_{A_*} = \mbox{max}(\chi_A)$. If $K\gg 1$, to avoid negative $\gamma_{A_*}$, we would have to assume $\chi_B^2\Lambda_B/(1+\zeta_B^2\Lambda_B)\ll 1$ for most $B\neq A_*$. If, for such $B$, $\zeta_B^2\Lambda_B \gsim 1$, then we have $\chi_B^2/\zeta_B^2\ll 1$ and most pair-wise cluster correlations are small. If, instead, $\zeta_B^2\Lambda_B \ll 1$, then $\chi_B^2\Lambda_B\ll 1$, and $\zeta_B\Omega_i\ll 1$ and $|\chi_B|\Omega_i\ll 1$ for $i\in J(B)$, so, nonsensically, the within-cluster pair-wise stock correlations ${\widehat \Gamma}_{ij} = \Omega_i\Omega_j\phi_{BB} \ll 1$, $i\neq j$, $i,j\in J(B)$.\footnote{\, Further, if all $\zeta_A^2\Lambda_A \ll 1$, both within-cluster and inter-cluster pair-wise stock correlations ${\widehat \Gamma}_{ij} = \Omega_i\Omega_j\phi_{G(i),G(j)} \ll 1$, $i\neq j$, unless all $|\chi_A| \gg \zeta_A$, i.e., unless, nonsensically, all clusters are almost 100\% (anti-)correlated. That is, in the (unrealistic) $\zeta_A^2\Lambda_A \ll 1$ limit $\Gamma_{ij}$ is almost diagonal. \label{fn.Lambda.zeta}}

\subsubsection{Cluster Weights}

{}So, how can (should) we compute $w_i$ such that they are nonnegative? We can get insight by looking at (\ref{gamma.A}). Let us, ad hoc, take all $\chi_A$ to be uniform: $\chi_A \equiv \chi$. Then we have
\begin{equation}\label{gamma.A.pos}
 \gamma_A = {\kappa^{-1}\over{1+\zeta_A^2\Lambda_A}}
\end{equation}
and all $\gamma_A > 0$. So, up to an immaterial overall normalization factor, this is the same result as what we get if we assume that $\phi_{AB}$ is diagonal, in which case we have $\gamma_A = 1/(1+\phi_{AA}^2\Lambda_A)$. However, here $\phi_{AB}$ is not diagonal and in (\ref{gamma.A.pos}) we have the {specific variance} $\zeta_A^2$ in the denominator, not the total variances $\phi_{AA}$. This is because we have effectively dropped the ``market mode" (i.e., the factor risk) in $\phi_{AB}$. What remains is the specific risk. So, the question is, what is the interpretation of (\ref{gamma.A.pos})?

{}Let us look at each cluster independently from other clusters. We can construct the benchmark for the universe of stocks corresponding to each cluster using the 1-factor model approach. These weights are given by (see (\ref{w.1-factor.beta}))
\begin{eqnarray}\label{w.A}
 &&[w(A)]_i = \eta_A~{\beta_i\over\xi_i^2},~~~i\in J(A)\\
 &&\eta_A^{-1} = \sum_{j\in J(A)}{\beta_j^2\over\xi_j^2} = \Lambda_A
\end{eqnarray}
Now we can construct cluster returns, i.e., the returns of the $K$ benchmark portfolios corresponding to the $K$ clusters, via
\begin{equation}
 R_A = \sum_{i\in J(A)} [w(A)]_i~R_i
\end{equation}
Since (up to an overall normalization factor) $E_i = \beta_i$ are the expected returns for the stocks, the expected returns $E_A$ for the clusters are $E_A \equiv 1$. If we construct a ``global" benchmark portfolio made of all the clusters, the corresponding weights with which we combine the clusters are (normalized such that $\sum_{A=1}^K w_A = 1$)
\begin{eqnarray}
 &&w_A = \mu~{E_A\over\zeta_A^2} = \mu~{1\over\zeta_A^2}\\
 &&\mu^{-1} = \sum_{A=1}^K {E_A\over\zeta_A^2} = \sum_{A=1}^K {1\over\zeta_A^2}
\end{eqnarray}
The stock weights in this ``global" benchmark portfolio are given by
\begin{equation}
 w_i = w_A~[w(A)]_i = {\mu\over\zeta_A^2\Lambda_A}~{\beta_i\over\xi_i^2},~~~i\in J(A)
\end{equation}
This is precisely (\ref{w.cluster}) with $\gamma_A$ of the form (\ref{gamma.A.pos}) and $\sigma_F^2 = \mu~\kappa$ (see (\ref{gamma.Lambda})) in the limit where $\zeta_A^2~\Lambda_A \gg 1$. So, the question is, what is the meaning of the extra 1 in (\ref{gamma.A.pos})?

{}To understand this, let us consider the opposite limit, where $\zeta_A^2~\Lambda_A \ll 1$. In this limit pair-wise stock correlations are small (see fn. \ref{fn.Lambda.zeta}).\footnote{\, Both within- and inter-cluster pair-wise stock correlations are small unless all $|\chi_A| \gg \zeta_A$.} This means that the total risk approximately equals the specific risk and the factor model covariance matrix is approximately diagonal. So, in this limit the effect of the clusters is negligible and we should recover our result for the 1-factor model (\ref{w.1-factor.beta}). And this is precisely what happens in the $\zeta_A^2~\Lambda_A \ll 1$ limit in (\ref{gamma.Lambda}) as $\gamma_A \approx 1/\kappa$ and is independent of $A$, so (\ref{w.cluster}) correctly reduces to (\ref{w.1-factor.beta}). For the intermediate values $\zeta_A^2~\Lambda_A \sim 1$, (\ref{gamma.Lambda}) smoothly interpolates between the two limits. This is what the full optimization gives, which balances the stock-specific risk and the factor risk. The only ``loose end" is that in arriving at (\ref{gamma.Lambda}) we assumed uniform $\chi_A \equiv \chi$, which results in $\gamma_A$ that na\"{\i}vely might appear as independent of $\chi$.\footnote{\, Up to an immaterial overall normalization, that is: $\kappa$ explicitly depends on $\chi$.} However, $\Lambda_A$ depends on ${\widehat \xi}_i^2 = 1 - \Omega_i^2~\phi_{AA} = 1 - \Omega_i^2\left(\zeta_A^2 + \chi^2\right)$, $i\in J(A)$. Furthermore, $\zeta_A^2$ depends on $\chi$ (which is the factor loading in the 1-factor model (\ref{1-factor.phi}) for $\phi_{AB}$) via a computation involving the time-series of factor returns (see \cite{Het}, \cite{HetPlus}). So, the aforesaid ``loose end" is this: why are $\chi_A\equiv \chi$ uniform?

{}This is because the cluster expected returns are uniform: $E_A\equiv 1$ (up to an overall normalization factor). Therefore, the corresponding factor betas $\beta_A$ are uniform: $\beta_A \equiv b$. So, following our discussion in Subsection \ref{sub.back.1-factor}, the factor loading $\Omega_A$ in the 1-factor model for the {\em correlation} matrix $\psi_{AB} = \phi_{AB}/\sigma_A\sigma_B$ (where $\sigma_A^2 = \phi_{AA}$) is given by $\Omega_A = \beta_A/\sigma_A$, while the corresponding factor loading in the {\em covariance} matrix $\chi_A = \sigma_A~\Omega_A$ is simply $\chi_A = \beta_A \equiv b$. I.e., $\chi_A$ are uniform, $\chi_A\equiv \chi$, and $\chi$ is identified with $b$. Note that while our discussion here na\"{\i}vely may appear a bit ``cavalier" w.r.t. the normalizations of $\beta_A$ and $\chi_A$, it is not. This is because, in a 1-factor model, the factor loading $\chi_A$ subsumes the $1\times 1$ factor covariance matrix, call it $\varphi$. So, the factor model (\ref{1-factor.phi}) actually reads:
\begin{equation}
 \phi_{AB} = \zeta_A^2~\delta_{AB} + \chi^\prime_A~\varphi~\chi^\prime_B
\end{equation}
Here $\chi_A^\prime$ is the raw (unnormalized) factor loading and $\chi_A = \sqrt{\varphi}~\chi^\prime_A$. So, we can identify $\chi_A^\prime$ with $\beta_A \equiv b$, which can be normalized arbitrarily, and this normalization is then subsumed in $\chi_A$ via $\varphi$, which is computed based on the time-series of factor returns and depends on $\chi_A^\prime$. The end result is that our $\chi_A$ are uniform: $\chi_A\equiv \chi$.

\subsubsection{A Generalization}

{}Above we discussed a 1-factor model for the cluster factor covariance matrix $\phi_{AB}$. However, we can generalize our result to a multifactor model for $\phi_{AB}$ where the $K$ clusters labeled by $A$ can be grouped into further $F$ clusters (typically, $F \ll K$), which we will label by $a$, $a = 1,\dots,F$. This naturally arises in binary fundamental industry classifications.\footnote{\, This structure also arises in statistical industry classifications \cite{StatIC}.} E.g., in BICS (see above) at the most granular level we have sub-industries, which are grouped into industries, which themselves are grouped into sectors. This chain can be thought of as ending with the final grouping into a single cluster corresponding to the (broad) ``market". So, the 1-factor model above could, e.g., describe BICS sectors, and the single factor loading corresponds to the ``market". Alternatively, we could take sub-industries or industries and build 1-factor models for them going straight to the ``market" level. However, instead, we can build multifactor models, e.g., $A$ labels sub-industries and $a$ labels industries, or $A$ labels subs-industries and $a$ labels sectors, etc. So, the idea here\footnote{\, This is the original nested ``Russian-doll" embedding of \cite{RD} used in \cite{Het} and \cite{HetPlus}.} is that we have an $F$-factor model for $\phi_{AB}$:
\begin{eqnarray}
 &&\phi_{AB} = \zeta_A^2~\delta_{AB} + \sum_{a,b=1}^F \chi_{Aa}~\varphi_{ab}~\chi_{Bb}\\
 &&\chi_{Aa} = \chi_A~\delta_{S(A), a}\\
 &&S:\{1,\dots,K\}\mapsto\{1,\dots,F\}
\end{eqnarray}
Here: $\chi_{Aa}$ is a $K\times F$ factor loadings matrix; $\varphi_{ab}$ is an $F\times F$ factor covariance matrix; the $K$-vector $\chi_A$ is a priori arbitrary; $S$ maps the $K$ ``sub-clusters" labeled by $A$ to the $F$ ``clusters" labeled by $a$, $a=1,\dots,F$; each cluster contains one and only one
sub-cluster; $J^\prime(a) = \{A|S(A) = a\}$ is the set of sub-clusters that belong to the cluster labeled by $a$. Straightforward algebra gives:
\begin{eqnarray}
 &&Q_{AB}^{-1} = {1\over\nu_A^2}~\delta_{AB} + {\chi_A\over\nu_A^2\zeta_A^2}~{\chi_B\over\nu_B^2\zeta_B^2}~\kappa_{S(A), S(B)}^{-1}\\
 &&\nu_A^2 = {1\over\zeta_A^2} + \Lambda_A \\
 &&\kappa_{ab} = \varphi^{-1}_{ab} + \delta_{ab}~\sum_{A\in J^\prime(a)} {\chi_A^2 \Lambda_A\over{1+\zeta_A^2\Lambda_A}}
\end{eqnarray}
So, we have
\begin{equation}\label{gamma.A.multi}
 \gamma_A = {1\over{1+\zeta_A^2\Lambda_A}}\left[ 1 - \chi_A~\sum_{B=1}^K {\chi_B\Lambda_B \over{1+\zeta_B^2\Lambda_B}}~\kappa_{S(A), S(B)}^{-1}\right]
\end{equation}
Following our logic above, we must take uniform $\chi_A \equiv \chi$. However, unlike the 1-factor case, where $\kappa$ was a number, here instead we have a matrix, $\kappa_{ab}$, which depends on the details of $\varphi_{ab}$. Consider a 1-factor model:
\begin{equation}
 \varphi_{ab} = \rho_a^2~\delta_{ab} + \omega_a~\omega_b
\end{equation}
For the same reason as why $\chi_A \equiv\chi$ are uniform, we must take uniform $\omega_a \equiv \omega$.\footnote{\, Thus, we have $K$ sub-clusters grouped into $F$ clusters. Similarly to (\ref{w.A}), we can construct the benchmark for the universe of sub-clusters corresponding to each cluster (recall that $E_A\equiv 1$): $[w(a)]_A = \eta_a/\zeta_A^2$, $A\in J^\prime(a)$, where $\eta^{-1}_a = \sum_{A\in J^\prime(a)}\zeta_A^{-2}$. We can compute the $F$ cluster returns $R_a$ using the $K$ sub-cluster returns: $R_a = \sum_{A\in J^\prime(a)}[w(a)]_A R_A$. Then the cluster expected returns $E_a \equiv 1$. Hence uniform $\omega_a$. Again, the foregoing holds up to immaterial overall normalizations.}

{}With uniform $\omega_a$, straightforward algebra yields the following simple result:
\begin{eqnarray}\label{gamma.A.3}
 && \gamma_A = {1\over{1+\zeta_A^2\Lambda_A}}~{1\over{1+\rho_{S(A)}^2\lambda_{S(A)}}}~{1\over{1+\tau}}\\
 && \lambda_a = \chi^2~\sum_{A\in J^\prime(a)} {\Lambda_A \over{1+\zeta_A^2\Lambda_A}}\\
 && \tau = \omega^2~\sum_{a=1}^F {\lambda_a \over{1+\rho_a^2\lambda_a}}
\end{eqnarray}
Note that the factorization in (\ref{gamma.A.3}) occurs precisely because $\chi_A$ and $\omega_a$ are uniform.\footnote{\, Let us emphasize that, unlike for stocks, uniform factor loadings are reasonable for clusters (e.g., industries, sectors, etc.). This is because clusters are diversified stock portfolios with much less skewed volatilities than stocks. Therefore, in this regard, very small clusters should be avoided.}

\subsection{General ``Russian-doll" Embedding}

{}Above we considered a 2-level clustering scheme. It is now evident how to generalize it to any $P$-level clustering scheme. We have the following sequence: Stocks (Level-0) $\rightarrow$ Level-1 Clusters $\rightarrow$ Level-2 Clusters $\rightarrow$ \dots $\rightarrow$ Level-$P$ Clusters $\rightarrow$ ``Market" (Level-$(P+1)$). Here ``Market" means the entire universe of $N$ stocks and can be thought of as the final single cluster in the above sequence.\footnote{\, Also, the last leg in the above sequence ``Level-$P$ Clusters $\rightarrow$ ``Market" (Level-$(P+1)$)" can be treated as optional and omitted, if so desired (see below).} Thus, BICS is a 3-level industry classification ($P=3$), where Level-1 Clusters = BICS Sub-industries, Level-2 Clusters = BICS Industries, and Level-3 Clusters = BICS Sectors.\footnote{\, Note that GICS (see above) has $P=4$ levels.} We have the following nested ``Russian-doll" risk model construction (here $\ell=1,\dots,P$):
\begin{eqnarray}
 \Gamma^{(\ell)}_{A^{(\ell)},B^{(\ell)}} &=& \left[\zeta^{(\ell)}_{A^{(\ell)}}\right]^2~\delta_{A^{(\ell)},B^{(\ell)}} + \nonumber \\
 &+& \sum_{A^{(\ell+1)},B^{(\ell+1)}=1}^{K^{(\ell+1)}} \Omega^{(\ell)}_{A^{(\ell)},A^{(\ell+1)}}~\Gamma^{(\ell+1)}_{A^{(\ell+1)},B^{(\ell+1)}}~\Omega^{(\ell)}_{B^{(\ell)},B^{(\ell+1)}}\label{RusDoll}
\end{eqnarray}
Here: $A^{(0)}, B^{(0)} = 1,\dots, N$ label stocks (i.e., they are the indices $i,j=1,\dots,N$ in the notations above); $\Gamma^{(0)}_{ij} = \Gamma_{ij}$ is the factor model {\em covariance} matrix for stocks (and $[\zeta^{(0)}_i]^2 = \xi_i^2$ are the corresponding specific variances); $A^{(\ell)}, B^{(\ell)} = 1,\dots, K^{(\ell)}$, $\ell = 1,\dots,P$, label the Level-$\ell$ Clusters; $\Gamma^{(\ell)}_{A^{(\ell)},B^{(\ell)}}$, $\ell=1,\dots,P$, are the factor covariance matrices corresponding to the Level-$\ell$ Clusters; at Level-$(P+1)$ we have $A^{(P+1)} = B^{(P+1)} = 1$ (i.e., these indices take only one value corresponding to the ``Market", so we have $K^{(P+1)} = 1$), and we can either have $[\zeta^{(P+1)}_1]^2 = \Gamma^{(P+1)}_{11} > 0$ (so $\Gamma^{(P)}_{A^{(P)},B^{(P)}}$ is a 1-factor model), or we can set $[\zeta^{(P+1)}_1]^2 = \Gamma^{(P+1)}_{11} = 0$ (so $\Gamma^{(P)}_{A^{(P)},B^{(P)}}$ is diagonal, i.e., a ``0-factor model"); finally, for the factor loadings $\Omega^{(\ell)}_{A^{(\ell)},A^{(\ell+1)}}$ we have (here, as above, $\beta_i$ are the stock betas):
\begin{eqnarray}
 && \Omega^{(0)}_{i,A^{(1)}} = \beta_i~\delta_{G^{(0)}(i), A^{(1)}}\\
 && \Omega^{(\ell)}_{A^{(\ell)},A^{(\ell+1)}} = \chi^{(\ell)}~\delta_{G^{(\ell)}(A^{(\ell)}), A^{(\ell+1)}},~~~\ell=1,\dots,P\\
 && G^{(\ell)}: \{1,\dots,K^{(\ell)}\} \mapsto \{1,\dots,K^{(\ell+1)}\},~~~\ell=0,1,\dots,P
\end{eqnarray}
Here: $G^{(\ell)}$ is a map from the Level-$\ell$ Clusters to the Level-$(\ell+1)$ Clusters, $\ell=0,1,\dots,P$; ``Level-0 Clusters" = stocks; $K^{(0)} = N$; at Level-$(P+1)$ we have the ``Market" (a single ``cluster"). The benchmark portfolio weights are then given by:
\begin{eqnarray}\label{w.multi.gen}
 && w_i = \sigma_F^2~{\beta_i\over\xi_i^2}~\gamma_{G^{(0)}(i)}\\
 &&\gamma_{A^{(1)}} = \prod_{\ell=1}^{P+1} \left(1 + \left[\zeta^{(\ell)}_{{\cal F}^{(\ell)}(A^{(1)})}\right]^2 \Lambda^{(\ell)}_{{\cal F}^{(\ell)}(A^{(1)})}\right)^{-1}\label{gamma.prod}\\
 &&{\cal F}^{(1)}(A^{(1)}) = A^{(1)}\\
 &&{\cal F}^{(\ell+1)}(A^{(1)}) = G^{(\ell)}({\cal F}^{(\ell)}(A^{(1)})) = G^{(\ell)}(G^{(\ell-1)}(\dots G^{(1)}(A^{(1)})\dots))\\
 && \Lambda^{(1)}_{A^{(1)}} = \sum_{j \in J^{(0)}(A^{(1)})} {\beta_j^2\over\xi_j^2}\\
 && \Lambda^{(\ell+1)}_{A^{(\ell+1)}} = \left[\chi^{(\ell)}\right]^2 \sum_{A^{(\ell)}\in J^{(\ell)}(A^{(\ell+1)})} \Lambda^{(\ell)}_{A^{(\ell)}} \left(1 + \left[\zeta^{(\ell)}_{A^{(\ell)}}\right]^2 \Lambda^{(\ell)}_{A^{(\ell)}}\right)^{-1}\label{Lambda.multi}\\
 && \sigma_F^{-2} = \sum_{A^{(1)}=1}^{K^{(1)}} \Lambda^{(1)}_{A^{(1)}}~\gamma_{A^{(1)}}
\end{eqnarray}
Here: $\ell = 1,\dots,P$ in (\ref{Lambda.multi}); the sets $J^{(\ell)}(A^{(\ell+1)}) = \{A^{(\ell)}| G^{(\ell)}(A^{(\ell)}) = A^{(\ell+1)}\}$; and ${\cal F}^{(P+1)}(A^{(1)})=1$.\footnote{\, Further, note that the factor in the product (\ref{gamma.prod}) corresponding to $\ell=P+1$ is actually independent of $A^{(1)}$, and it is equal 1 if $\zeta^{(P+1)}_1 = 0$ (i.e., if $\Gamma^{(P)}_{A^{(P)},B^{(P)}}$ is a ``0-factor model").} The benchmark weights (\ref{w.multi.gen}) comprise one of our main results.

\subsection{Risk Model Construction}

{}Assuming all $\beta_i > 0$ (and all $\chi^{(\ell)}>0$), it is not all that surprising that we can construct all-positive $w_i$ as the matrix $\Gamma_{ij}$ has all positive elements. Thus, pursuant to the Perron-Frobenius theorem \cite{Perron}, \cite{Frobenius}, all $V^{(1)}_i > 0$ (or can be chosen to be such as the signs of all $V^{(1)}_i$ can always be flipped simultaneously), where $V^{(1)}_i$ is the first principal component of $\Gamma_{ij}$. However, (\ref{w.multi.gen}) is not based on principal components of some random positive covariance (or correlation) matrix $\Gamma_{ij}$. Instead, here we construct a {\em non-random, meaningful} $\Gamma_{ij}$ for ``arbitrary" $\beta_i>0$.

{}Now, this is where some qualifications are in order. It is clear that $\beta_i$ must be skewed similarly to $\sigma_i$, where $\sigma_i^2 = \Gamma_{ii} = C_{ii}$ are sample variances for stocks. I.e., ${\widehat\beta}_i = \beta_i/\sigma_i$ is the quantity that is expected not to be skewed. Otherwise, within the same Level-1 Cluster labeled by $A^{(1)}$ stocks with large $\sigma_i$, $i\in J^{(0)}(A^{(1)})$, would have small correlations with other stocks. So, at Level-0 we have the following factor model ${\widehat \Gamma}_{ij}$ for the {\em correlation} matrix
\begin{eqnarray}
 &&\Gamma_{ij} = \sigma_i\sigma_j{\widehat \Gamma}_{ij}\\
 &&{\widehat \Gamma}_{ij} =  {\widehat \xi}_i^2~\delta_{ij} + {\widehat \beta}_i~\Gamma^{(1)}_{G^{(0)}(i), G^{(0)}(j)}~{\widehat \beta}_j\\
 &&{\widehat \xi}_i^2 = \xi_i^2/\sigma^2_i\\
 &&{\widehat \xi}_i^2 + {\widehat \beta}_i^2~\Gamma^{(1)}_{G^{(0)}(i), G^{(0)}(i)} \equiv 1\label{xi.cond}
\end{eqnarray}
As discussed in \cite{Het}, \cite{HetPlus}, it is the condition (\ref{xi.cond}) that is difficult to satisfy. In particular, for generic values of ${\widehat \beta}_i$ we would have to use the method of Section 4 of \cite{HetPlus}, whereby we have some trial values ${\widehat \beta}_i^\prime$ and the actual values ${\widehat \beta}_i$ are related to ${\widehat \beta}_i^\prime$ via a highly nonlinear combination of ${\widehat \beta}_i^\prime$ and the sample correlation matrix $\Psi_{ij}$. It is then impracticable to detangle ${\widehat \beta}_i^\prime$ from the desired values ${\widehat \beta}_i$. To avoid complications with such nonlinearities, we can use the heterotic construction of \cite{Het}, \cite{HetPlus}, where ${\widehat\beta}_i$, $i\in J^{(0)}(A^{(1)})$, are given by the first principal component of the square block $\Psi_{ij}$, $i,j\in J^{(0)}(A^{(1)})$. However, these principal components are not guaranteed to be all positive. This can be overcome by deforming each block such that all correlations therein are positive. This is doable but somewhat convoluted and there is no unique way of doing this. At the end we would have just a single choice of $\beta_i$ -- subject to variability due to the choice of the deformation, that is.\footnote{\, To be clear, the method of Section 4 of \cite{HetPlus} is perfectly adequate if we wish to construct a factor model, e.g., for optimization purposes, and need not worry about the precise values of the factor loadings (i.e., the fact that the trial and actual loadings are not the same). However, the problem at hand is different, which is to construct a benchmark portfolio for given betas, and the actual factor loadings must coincide with these betas by construction.}

{}Another possibility is to take\footnote{\, Note that equivalently we can take ${\widehat \beta}_i = b_{G^{(0)}(i)}$, where $b_{A^{(1)}} > 0$ and are otherwise arbitrary. This is because any rescaling ${\widehat \beta}_i \rightarrow  {\widehat \beta}_i~b_{G^{(0)}(i)}$ is simply absorbed by the corresponding rescaling of the factor covariance matrix $\Gamma^{(1)}_{A^{(1)}, B^{(1)}} \rightarrow \Gamma^{(1)}_{A^{(1)}, B^{(1)}}/b_{A^{(1)}}b_{B^{(1)}}$, so the factor model is unaffected.} ${\widehat\beta}_i\equiv 1$ and use the binary construction in Subsection 3.2 of \cite{HetPlus}. In this case we do not need to deform the sample correlation matrix. However, here too we have a single choice of $\beta_i$.\footnote{\, Also, this method would not work for the Level-$\ell$ Clusters, $\ell \geq 1$ (see below).}

\subsubsection{Computing Specific Risk}\label{sub.spec.risk}

{}Happily, precisely because for each block we are dealing with a 1-factor model, we can use another, simple method to satisfy (\ref{xi.cond}). Consider a symmetric $M\times M$ matrix $X_{\alpha\beta}$, $\alpha,\beta = 1,\dots, M$. For our purposes here it will suffice to assume that $X_{\alpha\beta}$ is semi-positive definite (as here we are interested in cases where $X_{\alpha\beta}$ is a covariance or correlation matrix). Suppose we wish to model it via a 1-factor model:
\begin{equation}
 Y_{\alpha\beta} = a_\alpha^2~\delta_{\alpha\beta} + b_\alpha~\vartheta~b_\beta
\end{equation}
subject to reproducing the diagonal elements:
\begin{eqnarray}
 && Y_{\alpha\alpha} = X_{\alpha\alpha} = \varsigma_\alpha^2\\
 && a_\alpha^2 + \vartheta~b_\alpha^2 = \varsigma_\alpha^2\label{ab}
\end{eqnarray}
So, we need to fit the unknown $\vartheta$ given the values of $b_\alpha$ and $X_{\alpha\beta}$. We can do this as follows. First, let us define $z_{min}$ and $z_{max}$ such that for all values of $\alpha$ we have
\begin{equation}
 z_{min}~\varsigma_\alpha \leq a_\alpha \leq z_{max}~\varsigma_\alpha
\end{equation}
I.e., $z_{min}$ and $z_{max}$ define the minimum and maximum allowed values of the fraction of the total standard deviation $\varsigma_\alpha$ attributable to the specific risk $a_\alpha$. (E.g., we can set $z_{min} = 0.1$ and $z_{max} = 0.9$.) This means that
\begin{eqnarray}\label{vartheta}
 &&\vartheta_{min} \leq \vartheta \leq \vartheta_{max}\\
 &&\vartheta_{min} = \left(1-z_{max}^2\right) / \mbox{min}({\widehat b}_\alpha^2)\label{theta.min}\\
 &&\vartheta_{max} = \left(1-z_{min}^2\right) / \mbox{max}({\widehat b}_\alpha^2)\label{theta.max}\\
 &&{\widehat b_\alpha} = b_\alpha / \varsigma_\alpha
\end{eqnarray}
So, given $\varsigma_\alpha$, the values of $b_\alpha$ cannot be arbitrary but must be such that $\vartheta_{min}\leq\vartheta_{max}$ (see Appendix \ref{app.A} for how $\vartheta_{min}>\vartheta_{max}$ cases are dealt with). Next, we can find the value of $\vartheta$ which provides the least-squares fit of the {\em off-diagonal} elements of
\begin{equation}
 {\widehat Y}_{\alpha\beta} = Y_{\alpha\beta}/\varsigma_\alpha\varsigma_\beta
\end{equation}
into those of ${\widehat X}_{\alpha\beta} = X_{\alpha\beta}/\varsigma_\alpha\varsigma_\beta $ (note that the diagonal elements ${\widehat Y}_{\alpha\alpha}\equiv 1$ need not be fit as they are fixed via (\ref{ab})):
\begin{equation}
 \sum_{\alpha,\beta=1;~\alpha\neq\beta}^M\left[{\widehat X}_{\alpha\beta} - {\widehat b}_\alpha~\vartheta~{\widehat b}_\beta\right]^2\rightarrow \mbox{min}
\end{equation}
subject to (\ref{vartheta}). So, we have
\begin{eqnarray}\label{opt.vartheta}
 &&\vartheta = \mbox{min}(\mbox{max}(\vartheta_*, \vartheta_{min}), \vartheta_{max})\\
 &&\vartheta_* = {{\sum_{\alpha,\beta=1;~\alpha\neq\beta}^M {\widehat b}_\alpha~{\widehat X}_{\alpha\beta}~{\widehat b}_\beta}\over{\sum_{\alpha,\beta=1;~\alpha\neq\beta}^M {\widehat b}_\alpha^2~{\widehat b}_\beta^2}}
\end{eqnarray}
Note that in our context here ${\widehat X}_{\alpha\beta}$ is a sample correlation matrix, so $|{\widehat X}_{\alpha\beta}|\leq 1$, in fact, $|{\widehat X}_{\alpha\beta}| < 1$ for $\alpha\neq\beta$. Assuming ${\widehat b}_\alpha$ are tightly distributed, we can expect $\vartheta_*$ to be somewhere between $\vartheta_{min}$ and $\vartheta_{max}$ (as opposed to saturating these bounds).

\subsubsection{Application to ``Russian-doll" Embedding}

{}Given $X_{\alpha\beta}$, $b_\alpha$, $z_{min}$ and $z_{max}$, let $\theta(X_{\alpha\beta}, b_\alpha, z_{min}, z_{max})$ denote the value of $\vartheta$ given by (\ref{opt.vartheta}). Then we have the following procedure for computing the specific risks and factor covariance matrices in the nested ``Russian-doll" embedding described above:\footnote{\, Below we suppress the $z_{min}, z_{max}$ arguments. In fact, they can be $\ell$-dependent, if so desired.}
\begin{eqnarray}
 &&X^{(0)}_{ij} = C_{ij}\\
 &&b^{(0)}_i = \beta_i\\
 &&b^{(\ell)}_{A^{(\ell)}} \equiv \chi^{(\ell)},~~~\ell = 1,\dots, P\\
 &&\Gamma^{(\ell+1)}_{A^{(\ell+1)},A^{(\ell+1)}} = \theta(X^{(\ell)}_{A^{(\ell)},B^{(\ell)}}, b^{(\ell)}_{A^{(\ell)}}),~~~
 A^{(\ell)},B^{(\ell)}\in J^{(\ell)}(A^{(\ell+1)})\label{vartheta.variances} \\
 &&X^{(\ell+1)}_{A^{(\ell+1)},B^{(\ell+1)}} = {\widetilde X}^{(\ell+1)}_{A^{(\ell+1)},B^{(\ell+1)}}~u^{(\ell+1)}_{A^{(\ell+1)}}~u^{(\ell+1)}_{B^{(\ell+1)}}\\
 &&u^{(\ell+1)}_{A^{(\ell+1)}} = \sqrt{\Gamma^{(\ell+1)}_{A^{(\ell+1)},A^{(\ell+1)}} \over {\widetilde X}^{(\ell+1)}_{A^{(\ell+1)},A^{(\ell+1)}}}\\
 &&{\widetilde X}^{(\ell+1)}_{A^{(\ell+1)},B^{(\ell+1)}} = \sum_{A^{(\ell)}\in J^{(\ell)}(A^{(\ell+1)})} \sum_{B^{(\ell)}\in J^{(\ell)}(B^{(\ell+1)})}
 X^{(\ell)}_{A^{(\ell)},B^{(\ell)}}~b^{(\ell)}_{A^{(\ell)}}~b^{(\ell)}_{B^{(\ell)}}\\
 &&\left[\zeta^{(\ell)}_{A^{(\ell)}}\right]^2 = X^{(\ell)}_{A^{(\ell)},A^{(\ell)}} - \left[b^{(\ell)}_{A^{(\ell)}}\right]^2 \Gamma^{(\ell+1)}_{G^{(\ell)}(A^{(\ell)}),G^{(\ell)}(A^{(\ell)})}
\end{eqnarray}
Here, as above, $C_{ij}$ is the sample covariance matrix of the stock returns. Also, note that the choice of $\chi^{(\ell)}$, $\ell = 1,\dots, P$, is immaterial. This procedure together with (\ref{RusDoll}) completely defines the risk model. All specific risks and factor covariance matrices are positive-definite by construction. And so are the benchmark weights (\ref{w.multi.gen}) for a range of values of $\beta_i$ so long as ${\widehat \beta}_i=\beta_i/\sigma_i$ are not skewed (see above).\footnote{\, Thus, if we, e.g., take $z_{min}=0.1$ and $z_{max}= 0.9$, then $\vartheta_{min} = 0.19/\mbox{min}({\widehat\beta}_i^2)$ and $\vartheta_{max} = 0.99/\mbox{max}({\widehat\beta}_i^2)$, so the allowed range of betas is $\mbox{max}({\widehat\beta}_i)/\mbox{min}({\widehat\beta}_i)\leq \sqrt{0.99/0.19}\approx 2.28$. Note that for the Level-$\ell$ Clusters, $\ell = 1,\dots, P$, generally it is reasonable to expect the factor covariances and specific risks to be of order 1 and nontrivial solutions for the fitted values of $\vartheta$ (i.e., $\Gamma^{(\ell+1)}_{A^{(\ell+1)},A^{(\ell+1)}}$ computed via (\ref{vartheta.variances})) to exist. Appendix \ref{app.A} deals with occasions when this does not hold.}

\subsubsection{Some Comments}

{}Here we can ask two questions. Why does the above construction make sense? And are there viable alternatives? Thus, above we construct the risk model in a rather specific way, grouping stocks into clusters and essentially building a 1-factor model within each cluster with the factor loadings given by the betas.\footnote{\, This is essentially the heterotic CAPM construction of \cite{HetPlus}, which is similar to the heterotic construction of \cite{Het}, except that in the latter the factor loadings are based on principal components, which are not necessarily all positive (see above).} Then we group these clusters into further clusters and repeat the procedure until we end up with a stable and positive-definite factor covariance matrix. The question we can ask is, can we take a more general risk model construction instead? Basically, there two separate issues here. The first issue is independent of the fact that we are dealing with a long-only portfolio and pertains to the fact that: i) higher-than-first principal components are intrinsically unstable out-of-sample; ii) standard style factors (see above) are poor proxies for modeling pair-wise correlations, so -- contrary to a common practice in commercial risk model offerings -- using them as factor loadings is highly suboptimal\footnote{\, Also, their number is limited and fails to compete with ubiquitous industry (cluster) factors.} (see \cite{HetPlus} for details); and iii) well-constructed fundamental\footnote{\, Statistical industry classifications are not as stable as fundamental industry classifications, but still sizably outperform models based on principal components \cite{StatIC}.} industry classifications are rather stable out-of-sample as companies rarely jump industries (let alone sectors). This is what justifies the above construction except for the choice of the factor loadings, which in this case are simply the betas. And this latter part of our construction is dictated by the fact that this choice is the only one that effectively removes the ``market mode". Other choices for factor loadings generically would lead to undesirable negative weights $w_i$.

{}In this regard, if we take a generic multifactor risk model for $\Gamma_{ij}$, which includes, e.g., style factors and/or principal components, some weights generically will be negative for any given choice of the betas. Let us emphasize that we can always work backwards, pick some positive weights $w_i$, and compute the corresponding betas $\beta_i$. However, it is unclear what any such portfolio represents. In contrast, in the above construction the meaning of the resultant benchmark portfolio is clear. Thus, assuming that the second term in the parentheses in (\ref{gamma.prod}) is dominant at each level $\ell$ (typically, this is expected to be a good approximation for large clusters) we have
\begin{eqnarray}
 && w_i \approx \eta~{\beta_i\over\xi_i^2\Lambda^{(1)}_{G^{(0)}(i)}}~\prod_{\ell=1}^{P} \left(\left[\zeta^{(\ell)}_{{\cal F}^{(\ell)}(G^{(0)}(i))}\right]^2 {\widetilde \Lambda}^{(\ell+1)}_{{\cal F}^{(\ell+1)}(G^{(0)}(i))}\right)^{-1}\\
 && \Lambda^{(1)}_{A^{(1)}} = \sum_{j \in J^{(0)}(A^{(1)})} {\beta_j^2\over\xi_j^2}\\
 && {\widetilde \Lambda}^{(\ell+1)}_{A^{(\ell+1)}} \approx \sum_{A^{(\ell)}\in J^{(\ell)}(A^{(\ell+1)})} \left[\zeta^{(\ell)}_{A^{(\ell)}}\right]^{-2}, ~~~\ell = 1,\dots,P\\
 &&\eta^{-1}\approx \sum_{A^{(1)}=1}^{K^{(1)}} \prod_{\ell=1}^{P} \left(\left[\zeta^{(\ell)}_{{\cal F}^{(\ell)}(A^{(1)})}\right]^2 {\widetilde \Lambda}^{(\ell + 1)}_{{\cal F}^{(\ell+1)}(A^{(1)})}\right)^{-1}
\end{eqnarray}
The interpretation of these weights (similarly to the example we discussed above) is clear: we suppress the weights by a product of {\em specific variances} at each level, with proper normalizations (such that at each level cluster betas are 1 up to immaterial overall normalization factors). Note that instead of specific variances we could use total variances, which would ``overcount" the ``market mode" risk, whereas specific variance exclude this factor risk. And this is just as well: the overall market exposure risk (i.e., the ``market mode" risk) is intrinsically present in long-only portfolios and should not be hedged against. Our construction neatly removes the ``market mode".

{}Finally, let us mention that a priori we could obtain a long-only portfolio for a given set of betas as follows. We can maximize the Sharpe ratio (\ref{Sharpe}) (where $E_i$ are related to the betas via (\ref{exp.ret})) subject to (\ref{norm.w.beta}) and lower bounds $w_i\geq w_i^{min}$, where $w_i^{min} \geq 0$. A priori here we can use a generic multifactor model covariance matrix $\Gamma_{ij}$ instead of the sample covariance matrix $C_{ij}$. However, for a generic $\Gamma_{ij}$ a large fraction of $w_i$ (in many cases, around 50\%) can turn out to saturate the lower bounds $w_i^{min}$. Such portfolios generically can be skewed and far from ``optimality".

\section{Beating the Market}\label{sec.4}

{}Now that we have a method for constructing benchmark portfolios, we can ask a different question. Suppose we have an ``alpha model", which forecasts expected returns $E_i$. Can we construct a long-only portfolio based on these returns? If all expected returns are positive, then we can treat them as betas, i.e., as in (\ref{exp.ret}), set $\beta_i = E_i/\gamma$ (where $\gamma$ is an immaterial normalization factor) and compute the weights $w_i$ via (\ref{w.multi.gen}). There are two independent issues with this approach. First, even if all $E_i\geq 0$, the distribution of $E_i/\sigma_i$ may be too wide, so using the method (\ref{opt.vartheta}) might be problematic.\footnote{\, Note that at the level of stocks (i.e., Level-0 in the above nomenclature), we can always find $\beta_i$ such that (\ref{opt.vartheta}) exists. Thus, we can take $\beta_i = \sigma_i$, so ${\widehat\beta}_i \equiv 1$. Then $\vartheta_* = {1\over N(N-1)}~\sum_{i,j=1;~i\neq j}^N \Psi_{ij}$, where $\Psi_{ij}$ is the sample correlation matrix. That is, $\vartheta_*$ in this case is the average pair-wise correlation, which typically is positive and well-within reasonably set bounds $\vartheta_{min}$ and $\vartheta_{max}$ for a given cluster (e.g., sub-industry, industry, sector) for a well-constructed industry classification.} Second, in practice, most alpha models will not have all nonnegative $E_i$, in fact, many $E_i$ (in many cases, around 50\%) may be negative.

{}So, how can we deal with this? Instead of constructing a long-only portfolio based on $E_i$ from scratch, we can follow a different, 2-step approach. Since we are building a long-only portfolio, we are exposed to market risk no matter what we do. So, we might as well identify a benchmark portfolio whose market exposure we are willing to live with.\footnote{\, This benchmark portfolio can (but need not) be constructed as in Section \ref{sec.3} (see below).} Then we can try to construct a long only portfolio that -- based on out-of-sample backtests -- can reasonably be {\em expected} (albeit, as with any forward-looking statements, not guaranteed) to outperform this benchmark portfolio. One way to construct this portfolio is to combine the benchmark portfolio with a {\em dollar-neutral} portfolio (such that the resultant portfolio is still long-only), where the dollar-neutral portfolio has a positive expected return and a low correlation with the benchmark portfolio. So, for the weights $w_i$ of our long-only portfolio, we have
\begin{eqnarray}
 && w_i = w^*_i + w^\prime_i\\
 &&\sum_{i=1}^N w_i^\prime = 0\label{dollar.neut}
\end{eqnarray}
where $w_i^\prime$ are the weights of the dollar-neutral portfolio such that
\begin{eqnarray}\label{bounds.1}
 && w^{min}_i \leq w^\prime_i\leq w_i^{max}\\
 && w^{min}_i \geq -w_i^*
\end{eqnarray}
and $w_i^* > 0$ are the benchmark weights. A priori we can take the upper bounds $w_i^{max} > 0$ to be infinity. We can also set tighter bounds, e.g., such that $w_i$ do no deviate from the benchmark portfolio by some percentage: $w_i^{min} = -z~w_i^*$, $w_i^{max} = z~w_i^*$, where, say, $0 < z < 1$. Other customizations/variations are possible.

{}The weights $w_i^\prime$ can be fixed in a variety of standard ways, e.g., via Sharpe ratio (or mean-variance -- see above) optimization, which is what we will assume for the sake of definiteness (albeit this is not critical here). Then, ignoring for a moment the bounds (\ref{bounds.1}) and the dollar-neutrality constraint (\ref{dollar.neut}),\footnote{\, A generic $\Gamma^{\prime-1}_{ij}$ includes the ``market mode", so unless the returns $E_i$ are contrivedly fine-tuned, the weights $w_i^\prime$, while not exactly dollar-neutral, generically are not expected to be highly skewed toward long or short positions. So, ignoring the dollar-neutrality condition is not detrimental. Without dollar-neutrality, assuming $\sum_{i=1}^N w_i^* = 1$, we no longer have $\sum_{i=1}^N w_i = 1$. However, this can be cured by simply rescaling $w_i$ (albeit this may move $w_i$ away from ``optimality"). Ignoring the bounds poses a bigger issue. However, we will incorporate both the dollar-neutrality and bounds in a moment. Ignoring them for now serves the purpose of developing an intuitive understanding.} we have
\begin{equation}
 w_i^\prime = \gamma^\prime~\sum_{j=1}^N \Gamma^{\prime-1}_{ij}~E_j
\end{equation}
Here $\gamma^\prime$ is a normalization coefficient (to be determined), and $\Gamma^{\prime-1}_{ij}$ is the inverse of $\Gamma^{\prime}_{ij}$, which is an $N\times N$ (typically, multifactor) model covariance matrix. Note that $\Gamma^{\prime}_{ij}$ need {\em not} be the same as $\Gamma^*_{ij}$, which denotes the multifactor model covariance matrix used in constructing the benchmark weights $w_i^*$. This is because $w_i^\prime$ is a long-short (dollar-neutral) portfolio, so we do not have the same kinds of restrictions on $\Gamma^{\prime}_{ij}$ as on $\Gamma^*_{ij}$. In fact, generally, we expect that $\Gamma^{\prime}_{ij}$ built using, say, the heterotic construction \cite{Het}, \cite{HetPlus} (which utilizes first principal components of the blocks of the sample correlation matrix corresponding to clusters) would work better than $\Gamma^*_{ij}$. Then, given some $\Gamma^{\prime}_{ij}$, what should $\gamma^\prime$ be?

{}Considering that $\Gamma^{\prime}_{ij}$ is the multifactor model covariance matrix we use for modeling risk of a generic portfolio, we can compute the expected Sharpe ratio of the combined portfolio as follows:
\begin{equation}
 S = {{\sum_{i=1}^N E_i~w_i}\over{\sqrt{\sum_{i,j=1}^N \Gamma^{\prime}_{ij}~w_i~w_j}}}
\end{equation}
Further, the expected correlation $\rho$ between the portfolios $w_i^*$ and $w_i^\prime$ is given by
\begin{equation}
 \rho = {1\over\sigma^*\sigma^\prime}~\sum_{i,j=1}^N \Gamma^{\prime}_{ij}~w^*_i~w^\prime_j = {\gamma^\prime E^*\over\sigma^*\sigma^\prime} = {E^*\over\sigma^*e^\prime}
\end{equation}
where $E^* $ is the expected return of the benchmark portfolio, and $\sigma^*$ and $\sigma^\prime$ are the expected volatilities of the $w_i^*$ and $w_i^\prime$ portfolios:
\begin{eqnarray}
 &&E^* = \sum_{i=1}^N E_i~w^*_i\\
 &&(\sigma^*)^2 = \sum_{i,j=1}^N \Gamma^{\prime}_{ij}~w^*_i~w^*_j\\
 &&(\sigma^\prime)^2 = \sum_{i,j=1}^N \Gamma^{\prime}_{ij}~w^\prime_i~w^\prime_j = (\gamma^\prime)^2~ (e^\prime)^2\\
 &&(e^\prime)^2 = \sum_{i,j=1}^N \Gamma^{\prime-1}_{ij}~E_i~E_j
\end{eqnarray}
So, for the Sharpe ratio $S$ as a function of $\gamma^\prime$ we have:
\begin{eqnarray}
 &&S(\gamma^\prime) = {{E^* + \gamma^\prime~(e^\prime)^2}\over \sigma(\gamma^\prime)} = {\partial\sigma(\gamma^\prime)\over\partial\gamma^\prime}\nonumber\\
 &&\sigma(\gamma^\prime) = \sqrt{(\sigma^*)^2 + 2~\gamma^\prime~ E^*+(\gamma^\prime)^2~(e^\prime)^2}
\end{eqnarray}
The Sharpe ratio is maximized when $\gamma^\prime\rightarrow \infty$.\footnote{\, Thus, we have $S(0) = E^*/\sigma^* = \rho~e^\prime$, and $S(\gamma^\prime\rightarrow\infty) = e^\prime > S(0)$. Also, $\partial S(\gamma^\prime)/\partial \gamma^\prime = (\sigma^*)^2(e^\prime)^2\left[1-\rho^2\right]/\sigma^3(\gamma^\prime) > 0$.} However, in this limit we do not have a long-only portfolio. Instead, we have a long-short portfolio, which, in fact, would be dollar-neutral had we incorporated the dollar-neutrality constraint. In actuality, we must impose the bounds (\ref{bounds.1}). Then, in the limit $\gamma^\prime\rightarrow 0$, we have the long-only portfolio $w_i^*$. As we increase $\gamma^\prime$, more and more bounds will be saturated. The bounds distort the $w_i^\prime$ portfolio away from ``optimality". Above some value $\gamma^\prime_{opt}$, the Sharpe ratio starts to fall off: $S(\gamma^\prime) < S(\gamma^\prime_{opt})$ for $\gamma^\prime > \gamma^\prime_{opt}$. We can fix $\gamma^\prime_{opt}$ via, e.g., the golden-section search \cite{Kiefer} and use it as the ``optimal value" of $\gamma^\prime$.

\subsection{Including Bounds and Constraints}

{}In practice, in the presence of the bounds (\ref{bounds.1}) it is easier to implement mean-variance optimization than maximizing the Sharpe ratio.\footnote{\, The two are not equivalent once bounds, costs, etc., are included, except only in one special case of establishing trades with linear costs \cite{MeanRev}.} In the mean-variance optimization, we maximize the objective function w.r.t. $w_i^\prime$ (for a fixed value of $\gamma^\prime$)
\begin{equation}\label{obj.fn}
 g(w_i^\prime,\gamma^\prime) = \sum_{i=1}^N E_i~w_i^\prime - {1\over\gamma^\prime}~\sum_{i,j=1}^N \Gamma^{\prime}_{ij}~w_i^\prime~w_j^\prime
\end{equation}
subject to the bounds (\ref{bounds.1}) and the dollar-neutrality constraint (\ref{dollar.neut}). This optimization can be performed in a standard way (see, e.g., \cite{MeanRev} for a detailed algorithm and \cite{Het} for the source code). In fact, we may wish to include additional linear constraints. Thus, the expected correlation $\rho$ generally is nonzero. So, the dollar-neutral portfolio $w_i^\prime$ has a nonzero expected beta with the benchmark portfolio. We may wish to make $\rho$ vanish. This can be achieved via the following linear homogeneous constrain on $w_i^\prime$:
\begin{eqnarray}\label{null.rho}
 &&\sum_{i=1}^N q_i~w_i^\prime = 0\\
 &&q_i = \sum_{j=1}^N \Gamma^{\prime}_{ij}~w_j^*
\end{eqnarray}
Alternatively, we may wish to make the $w_i^\prime$ portfolio simply orthogonal to the $w_i^*$ portfolio, which is achieved via the following constraint:
\begin{equation}
 \sum_{i=1}^N w_i^*~w_i^\prime = 0
\end{equation}
More generally, we can have $p$ constraints
\begin{equation}
 \sum_{i=1}^N Q_{ia}~w_i^\prime = 0,~~~a=1,\dots,p
\end{equation}
Here $Q_{ia}$ is an $N\times p$ matrix with linearly independent columns, one of which is the unit $N$-vector corresponding to the dollar-neutrality constraint. We can also include neutrality w.r.t. sectors, industries, etc., or some style factors, if so desired. Etc.

\subsubsection{A Comment}

{}We can achieve an approximately null expected correlation $\rho$ in a different way. For generic ``raw" expected returns $E_i$ we have nonzero $E^*$ and thus nonzero $\rho$. However, given an $N$-vector $E_i$, we can construct $E_i^\prime$ orthogonal to $w_i^*$, e.g., by regressing $E_i$ (with unit weights and no intercept) over $w_i^*$ and taking the residuals (i.e., $E_i^\prime = \epsilon_i$):
\begin{equation}
 \epsilon_i = E_i - w_i^*~{{\sum_{j=1}^N E_j~w_j^*}\over {\sum_{j=1}^N (w_j^*)^2}}
\end{equation}
More generally, we can use a weighted regression with weights $v_i$ (and no intercept):\footnote{\, Then the question is, what should the weights $v_i$ be? Basing them on volatilities $\sigma_i$ (or the corresponding specific risks) would make little sense as i) we are already optimizing the $w_i^\prime$ portfolio and ii) both $|E_i|$ and $w_i^*$ scale linearly with $\sigma_i$. So, we can simply take unit weights $v_i\equiv 1$, or base them on quantities that are independent of $\sigma_i$ or have milder dependence thereon (e.g., $\ln(\sigma_i)$).}
\begin{equation}
 \epsilon_i = v_i\left(E_i - w_i^*~{{\sum_{j=1}^N v_j~E_j~w_j^*}\over {\sum_{j=1}^N v_j~(w_j^*)^2}}\right)
\end{equation}
If in (\ref{obj.fn}) we substitute $\epsilon_i$ instead of $E_i$, then we {\em approximately} achieve (\ref{null.rho}).\footnote{\, If it were not for the distortion caused by the bounds (\ref{bounds.1}), (\ref{null.rho}) would be precisely satisfied.}

\section{Concluding Remarks}\label{sec.5}

{}Let us briefly conclude with some remarks. First, in the market outperformance strategy we discuss in Section \ref{sec.4}, the benchmark $w_i^*$ a priori can be any long-only portfolio (including S\&P 500, Russell 3000, etc.), and not just built using the method of Section \ref{sec.3}. However, market cap weighted portfolios and benchmark portfolios of Section \ref{sec.3} are expected to have sizable correlations. Intuitively this may appear to be evident as these are all long-only portfolios. However, this goes beyond such ``zeroth-approximation" intuition. Thus, the weights $w_i^*$ in the benchmarks of Section \ref{sec.3} scale as $\propto 1/\sigma_i$, and $-\ln(\sigma_i)$ and $\ln(M_i)$ ($M_i$ is the market cap) are highly correlated. The ``devil" then is in the details of the construction of Section \ref{sec.3}, which aims to capture the substructure in the stock returns corresponding to the multilevel industry classification (clustering). I.e., there is more information in $w_i^*$ than in $M_i$.

{}In our construction above, $\beta_i$, albeit not completely arbitrary, are not fixed. Recall from Section \ref{sec.3} that, up to an overall normalization factor, ${\widehat \beta}_i = \beta_i/\sigma_i$ are of order 1 with a tight distribution with a standard deviation also of order 1. So, what should these betas be? A simple answer is that there is no magic bullet here. We can simply pick them, backtest them out-of-sample and compare the results with those for other values.\footnote{\, This kind of ``sampling" can get computationally taxing quickly.} Or we can take the ``observed" values ${\widehat \beta}^{obs}_i$ based on some broad market index, calculate their median value ${\widehat \beta}^{median} = \mbox{median}({\widehat \beta}^{obs}_i)$, and then cap and floor the ``outliers" by\footnote{\, MAD = mean absolute deviation.} ${\widehat \beta}^{max} = {\widehat \beta}^{median} + \kappa^{max}~\mbox{MAD}({\widehat \beta}^{obs}_i)$ and ${\widehat \beta}^{min} = {\widehat \beta}^{median} - \kappa^{min}~\mbox{MAD}({\widehat \beta}^{obs}_i)$, where $\kappa^{max}\sim 1$ and $\kappa^{min}\sim 1$ (we can set $\kappa^{max} = \kappa^{min}$). Fixing ${\widehat \beta}_i$ this way is a ``roundabout" as it uses another (cap-weighted) benchmark. But then again, there are no ``first principles" that can fix ${\widehat \beta}_i$ uniquely.

{}Let us note that, while we can take ${\widehat \beta}_i\equiv 1$ in the construction of Section \ref{sec.3}, we cannot take $\beta_i \equiv 1$. However, as mentioned above, $\beta_i \equiv 1$ would make little sense to begin with. Indeed, using (\ref{def.beta}), we have
\begin{equation}
 \beta_i = {{\sigma_i~\rho_i}/\sigma_F}
\end{equation}
where, as before, $\sigma_F$ is the benchmark portfolio volatility, and $\rho_i$ is the sample correlation between the stock labeled by $i$ and the benchmark. Setting $\beta_i \equiv 1$ (again, up to an overall normalization factor) would imply that $\rho_i\propto 1/\sigma_i$. Similarly, in the factor model context of Section \ref{sec.3}, we would have that, within the same cluster (sector, industry, etc.) stocks with high volatilities are almost uncorrelated to stocks with low volatilities (and each other). And this is not what we observe empirically (see, e.g., \cite{Het}, \cite{HetPlus}). We must have $\beta_i\propto\sigma_i$.

{}To tie up the final ``lose end", the weights $w_i^\prime$ of the dollar-neutral portfolio in Section \ref{sec.4} can be computed using the bounded optimization code in Appendix C of \cite{Het}, to wit, the function {\tt{\small bopt.calc.opt()}}, whose arguments are: {\tt{\small ret}} is the $N$-vector of expected returns $E_i$; {\tt{\small load}} is the matrix of constraints $Q_{ia}$; {\tt{\small inv.cov}} is the inverse matrix $\Gamma^{\prime-1}_{ij}$; {\tt{\small upper}}, {\tt{\small lower}} are the bounds $w_i^{max}$, $w_i^{min}$.

\appendix
\section{R Source Code for Benchmark Weights}\label{app.A}

{}In this appendix we give the R (R Package for Statistical Computing, \url{http://www.r-project.org}) source code for computing the benchmark weights $w_i$ based on the algorithm of Section \ref{sec.3}. This code is essentially self-explanatory and straightforward as it simply follows the formulas therein. The function {\tt{\small qrm.benchmark(ret, ind, beta, mkt.fac = T, z.min = 0.1, z.max = 0.9)}} returns the weights $w_i$ normalized as in (\ref{norm.w.beta}). The input is as follows: {\tt{\small ret}} is an $N\times T$ matrix of returns $R_{is}$ (e.g., daily close-to-close returns), where $N$ is the number of tickers, $T$ is the number of observations in the time-series (e.g., the number of trading days), and the ordering of the dates is immaterial; ii) {\tt{\small ind}} is a list of length $P$, whose elements are populated by the binary matrices (with rows corresponding to tickers, so {\tt{\small dim(ind[[$\cdot$]])[1]}} is $N$) corresponding to the levels in the input binary industry classification in the order of decreasing granularity (for BICS {\tt{\small ind[[1]]}} is the $N\times K^{(1)}$ matrix $\delta_{G(i), A^{(1)}}$ (sub-industries), {\tt{\small ind[[2]]}} is the $N\times K^{(2)}$ matrix $\delta_{G^\prime(i), A^{(2)}}$ (industries), and {\tt{\small ind[[3]]}} is the $N\times K^{(3)}$ matrix $\delta_{G^{\prime\prime}(i), A^{(3)}}$ (sectors), where $G=G^{(0)}$ maps tickers to sub-industries, $G^\prime = G^{(0)}G^{(1)}$ maps tickers to industries, and $G^{\prime\prime} = G^{(0)}G^{(1)}G^{(2)}$  maps tickers to sectors); iii) {\tt{\small beta}} is the $N$-vector $\beta_i$; iv) {\tt{\small mkt.fac}}, where at the final step for {\tt{\small TRUE}} we have a single industry factor (``Market"), while for {\tt{\small FALSE}} the industry factors correspond to the least granular level in the industry classification (sectors for BICS); and v) {\tt{\small z.min}} and {\tt{\small z.max}} are $z_{min}$ and $z_{max}$ defined in Subsection \ref{sub.spec.risk}.

{}There are two small tweaks in the source code beyond what is in Section \ref{sec.3}. First, if Level-1 (in the nomenclature of Section \ref{sec.3}) is very granular, for a given universe of stocks we can have one or more Level-1 Clusters each containing only one stock (e.g., single-stock sub-industries can and do arise in BICS). Typically, for long-horizon long-only portfolios using such granularity can be overkill. However, just in case, the source code deals with these situations in the internal function {\tt{\small calc.theta()}}. Second, on occasion, it can happen that $\theta_{min} > \theta_{max}$ (these quantities are defined in (\ref{theta.min}) and (\ref{theta.max}) in Subsection \ref{sub.spec.risk}). This can happen when there are outliers with too low or too high variances. Now, here we can try to do all kinds of contrived and convoluted things. But there is a simple way of dealing with this situation. Note that by definition both $\theta_{min}$ and $\theta_{max}$ are positive. Violating $\theta_{max}$, i.e., having $\theta>\theta_{max}$, can -- unacceptably -- lead to negative specific variances (see Subsection \ref{sub.spec.risk}). On the other hand, violating $\theta_{min}$, i.e., having $\theta<\theta_{min}$ is not detrimental so long as $\theta > 0$. Indeed, in this case we just have small factor risk. So, $\theta_{min} > \theta_{max}$ can be simply dealt with by setting $\vartheta$ as in (\ref{opt.vartheta}) as opposed to
\begin{equation}
 \vartheta = \mbox{max}(\mbox{min}(\vartheta_*, \vartheta_{max}), \vartheta_{min})
\end{equation}
which is equivalent to (\ref{opt.vartheta}) when $\theta_{min} \leq \theta_{max}$, but not when $\theta_{min} > \theta_{max}$. So, the source code uses (\ref{opt.vartheta}); see the line {\tt{\small t <- min(max(t, t.min), t.max)}} in the function {\tt{\small calc.theta()}}. Note that if we set ${\widehat\beta}_i\equiv 1$, or more generally have $\mbox{max}({\widehat\beta}_i)/\mbox{min}({\widehat\beta}_i)\leq\sqrt{(1-z^2_{min})/(1-z^2_{max})}$, then we are guaranteed to have $\theta_{min} \leq \theta_{max}$ at Level-1. But $\theta_{min} > \theta_{max}$ can arise at less granular levels due to outliers in factor variances.\\
\\
{\tt{\small
\noindent qrm.benchmark <- function (ret, ind, beta, mkt.fac = T,\\
\indent z.min = 0.1, z.max = 0.9)\\
\{\\
\indent calc.load <- function(load, load1)\\
\indent \{\\
\indent \indent x <- colSums(load1)\\
\indent \indent load <- (t(load1) \%*\% load) / x\\
\indent \indent return(load)\\
\indent \}\\
\\
\indent calc.theta <- function(x, b, z.min, z.max)\\
\indent \{\\
\indent \indent if(length(x) == 1)\\
\indent \indent \indent return((1 - z.max\^{}2) * x / b\^{}2)\\
\indent \indent s <- sqrt(diag(x))\\
\indent \indent x <- t(x / s) / s\\
\indent \indent b <- b / s\\
\indent \indent t.min <- (1 - z.max\^{}2) / min(b\^{}2)\\
\indent \indent t.max <- (1 - z.min\^{}2) / max(b\^{}2)\\
\indent \indent x <- t(x * b) * b\\
\indent \indent x <- sum(x) - sum(diag(x))\\
\indent \indent b <- sum(b\^{}2)\^{}2 - sum(b\^{}4)\\
\indent \indent t <- x / b\\
\indent \indent t <- min(max(t, t.min), t.max)\\
\indent \indent return(t)\\
\indent \}\\
\\
\indent ind[[length(ind) + 1]] <- matrix(1, nrow(ind[[1]]), 1)\\
\indent x <- cov(t(ret))\\
\indent y <- list()\\
\indent v <- list()\\
\indent w <- b <- beta\\
\\
\indent for(lvl in 1:length(ind))\\
\indent \{\\
\indent \indent if(lvl > 1)\\
\indent \indent \{\\
\indent \indent \indent flm <- calc.load(ind[[lvl]], ind[[lvl - 1]])\\
\indent \indent \indent b <- rep(1, nrow(flm))\\
\indent \indent \}\\
\indent \indent else\\
\indent \indent \indent flm <- ind[[lvl]]\\
\\
\indent \indent G <- rep(0, k <- ncol(flm))\\
\indent \indent y1 <- rep(0, nrow(flm))\\
\indent \indent v1 <- rep(0, k)\\
\\
\indent \indent for(a in 1:k)\\
\indent \indent \{\\
\indent \indent \indent take <- flm[, a] == 1\\
\indent \indent \indent if(lvl == length(ind) \& !mkt.fac)\\
\indent \indent \indent \indent G[a] <- 0\\
\indent \indent \indent else\\
\indent \indent \indent \indent G[a] <- calc.theta(x[take, take], b[take],\\
\indent \indent \indent \indent \indent z.min = z.min, z.max = z.max)\\
\indent \indent \indent y1[take] <- diag(x)[take] - b[take]\^{}2 * G[a]\\
\indent \indent \indent if(lvl == 1)\\
\indent \indent \indent \indent v1[a] <- sum(b[take]\^{}2 / y1[take])\\
\indent \indent \indent else\\
\indent \indent \indent \indent v1[a] <- sum(v[[lvl - 1]][take] / \\
\indent \indent \indent \indent \indent (1 + y1[take] * v[[lvl - 1]][take]))\\
\indent \indent \}\\
\\
\indent \indent y[[lvl]] <- y1\\
\indent \indent v[[lvl]] <- v1\\
\indent \indent x1 <- t(flm) \%*\% x \%*\% flm\\
\indent \indent u <- sqrt(G / diag(x1))\\
\indent \indent x <- t(x1 * u) * u\\
\indent \}\\
\\
\indent w <- w / y[[1]]\\
\\
\indent for(lvl in 1:(length(ind) - 1))\\
\indent \{\\
\indent \indent for(a in 1:ncol(ind[[lvl]]))\\
\indent \indent \{\\
\indent \indent \indent take <- ind[[lvl]][, a] == 1\\
\indent \indent \indent w[take] <- w[take] / (1 + y[[lvl + 1]][a] * v[[lvl]][a])\\
\indent \indent \} \\
\indent \}\\
\\
\indent w <- w / sum(w * beta)\\
\indent return(w)\\
\}
}}

\section{DISCLAIMERS}\label{app.B}

{}Wherever the context so requires, the masculine gender includes the feminine and/or neuter, and the singular form includes the plural and {\em vice versa}. The author of this paper (``Author") and his affiliates including without limitation Quantigic$^\circledR$ Solutions LLC (``Author's Affiliates" or ``his Affiliates") make no implied or express warranties or any other representations whatsoever, including without limitation implied warranties of merchantability and fitness for a particular purpose, in connection with or with regard to the content of this paper including without limitation any code or algorithms contained herein (``Content").

{}The reader may use the Content solely at his/her/its own risk and the reader shall have no claims whatsoever against the Author or his Affiliates and the Author and his Affiliates shall have no liability whatsoever to the reader or any third party whatsoever for any loss, expense, opportunity cost, damages or any other adverse effects whatsoever relating to or arising from the use of the Content by the reader including without any limitation whatsoever: any direct, indirect, incidental, special, consequential or any other damages incurred by the reader, however caused and under any theory of liability; any loss of profit (whether incurred directly or indirectly), any loss of goodwill or reputation, any loss of data suffered, cost of procurement of substitute goods or services, or any other tangible or intangible loss; any reliance placed by the reader on the completeness, accuracy or existence of the Content or any other effect of using the Content; and any and all other adversities or negative effects the reader might encounter in using the Content irrespective of whether the Author or his Affiliates is or are or should have been aware of such adversities or negative effects.

{}The R code included in Appendix \ref{app.A} hereof is part of the copyrighted R code of Quantigic$^\circledR$ Solutions LLC and is provided herein with the express permission of Quantigic$^\circledR$ Solutions LLC. The copyright owner retains all rights, title and interest in and to its copyrighted source code included in Appendix \ref{app.A} hereof and any and all copyrights therefor.

\end{document}